\documentclass[10pt]{article}
\usepackage{eurosym}
\usepackage{amsmath}
\usepackage{amsfonts}
\usepackage{color}
\usepackage{latexsym,amssymb,enumerate,ulem,pifont,calc}

\makeatletter

\@addtoreset{equation}{section}
\makeatother

\begin{document}

\begin{titlepage}

\vskip 2.0 cm
\begin{center}  {\huge{\bf On Generalized Lemaitre-Tolman-Bondi Metric}} \\\vskip 0.5 cm {\Large{\bf Fractal  Matter at the end of Matter-Antimatter Recombination}}

\vskip 2.5 cm

{\large{\bf Sergio Cacciatori$^1$}, {\bf Alessio Marrani$^2$}, and {\bf Federico Re$^1$}}

\vskip 1.0 cm

$^1${\sl
DiSAT, Universit\`a dell'Insubria, Via Valleggio 11, Como, Italy\\
and INFN, sezione di Milano, Via Celoria 16, 20133, Milano, Italy\\
\texttt{sergio.cacciatori@uninsubria.it}, \texttt{fre@uninsubria.it}}\\

\vskip 0.5 cm

$^2${\sl Centro Studi e Ricerche Enrico Fermi, Via Panisperna 89A,
I-00184, Roma, Italy}\\
{\sl Dipartimento di Fisica
e Astronomia Galileo Galilei, Universit\`a di Padova,\\
and INFN, sezione di Padova, Via Marzolo 8, I-35131 Padova, Italy\\
\texttt{jazzphyzz@gmail.com}}\\

\vskip 1.5 cm

 \end{center}

 \vskip 6.0 cm

\begin{%
abstract}

Many recent researches have investigated the deviations from the
Friedmannian cosmological model, as well as their consequences on
unexplained cosmological phenomena, such as dark matter and the acceleration of the Universe. On the one hand, a first order perturbative study of matter
inhomogeneity returned a partial explanation of dark matter and dark energy,
as relativistic effects due to the retarded potentials of far objects. On
the other hand, the fractal cosmology, now modeled with a
Lemaitre-Tolman-Bondi (LTB) metric, results in distortions of the luminosity
distances of SNe Ia, explaining the acceleration as apparent. In this work
we extend the LTB metric to ancient times. The origin of the fractal
distribution of matter is explained as the matter remnant after the
matter-antimatter recombination epoch. We show that the evolution of such a
inhomogeneity necessarily requires a dynamical generalization of LTB, and we
propose a particular solution.
 \end{abstract}
\vspace{24pt} \end{titlepage}


\newpage \tableofcontents \newpage


\section{Introduction}

Recent years have witnessed various attempts to explain, at least partially,
dark matter and dark energy phenomena as general relativistic effects due to
the inhomogeneities in the distribution of matter, at large scales.

\subsection{Fractal cosmology}

An interesting approach to the study of matter inhomogeneity is based on its
description by fractal geometry. The hierarchical structure of galaxies and
cluster has allowed for the introduction of the so-called \textquotedblleft
cosmic fractal\textquotedblright , e.g. in \cite{fract}. Remarkably,
physical results can be obtained despite the lack of an exact form of the
matter density distribution, by focussing rather on its fractal properties,
such as the fractal dimension $D<3$. In \cite{fract}, the statistical
correlation between positions of galaxies has been used to estimate $D\cong
1.2$.

Since the usual Friedmann model assumes an homogeneous distribution of
matter, a fractal distribution generates different cosmological laws. The
solution of the Einstein Equations with a fractal source $\rho (\underline{x}%
)$ would be a formidable mathematical task, due to the absence of any
continuous symmetry, as well as for the singularity of the source: $\rho \in
D^{\prime }(\mathbb{R}^{3})$. However, one can obtain an approximated
solution by considering not the real fractal density $\rho $, but rather the
total mass inside any ball of radius $r$. Neglecting the \textquotedblleft
void bubbles\textquotedblright ~or the concentration of matter in different
directions, one can perform an \textquotedblleft
homogenization\textquotedblright \footnote{%
The approximation of $\bar{\rho}$ is isotropic, so maybe it should better be
called an \textquotedblleft isotropization\textquotedblright ~ procedure.}%
~of the fractal $\bar{\rho}(r)$, which enjoys a quite high symmetry,
depending only on $r$ and $t$. Then, the homogenized fractal allows for an
exact solution to the Einstein Equation, which is known as the
Lemaitre-Tolman-Bondi (LTB) metric, as derived e.g. in \cite{Ribeiro:2008rs}.

The use of the LTB metric could seemingly violate the Cosmological
Principle, as the center of coordinate reference frame plays a key role. But
one should bear in mind that the LTB metric is just an approximation of the
real metric, which has no center at all. In fact, the homogenization can be
performed with respect to any point within the fractal, taken as the center
of the balls, and this will result in the LTB metric to have such a point as
preferred. Since \textit{any} point can act as \textquotedblleft a
center\textquotedblright , the Cosmological Principle is restored. The
matter, and the number of galaxies, around a point $\underline{x}_{0}$ of
the fractal $\mathcal{F}$ must grow approximately as%
\begin{equation}
M(r):=M(B(\underline{x}_{0};r))\cong \Phi r^{D},
\end{equation}%
where $D$ is the fractal dimension and $\Phi $ is a \textquotedblleft
fractal density\textquotedblright . We will assume that $\Phi $ is the same
for all points $\underline{x}_{0}\in \mathcal{F}$, i.e. we consider an
\textquotedblleft homogeneous fractal\textquotedblright . Since $D<3$, the
homogenized density $\bar{\rho}(r)\propto r^{D-3}$ shows a singularity at $%
r=0$: this can be traced back to the singular nature of the fractal
distribution itself, so that it has an infinite (three-dimensional) density
at any point.

It is here worth remarking that the law $\propto r^{D}$ cannot be true for a
point $\underline{x}_{0}$ outside the fractal. Inside a void bubble, the
mass is zero. Thus, one can appreciate that there is actually no equivalence
between all points, but rather the fractal splits the points into two
different categories: the \textit{material points} $\underline{x}_{0}\in
\mathcal{F}$, and the \textit{void points}~outside. Indeed, in fractal
cosmology the complete Cosmological Principle cannot be assumed, as it gets
replaced by the \textit{Conditional Cosmological Principle}, which does not
refer to all observers, but only to the material ones \cite{Ribeiro:2008rs}.

Within this theoretical framework, various studies have been addressing the
differences between the FLRW and LTB cosmologies. E.g. in \cite%
{Cosmai:2018nvx}, the luminosity distances of SNe Ia were recalculated,
obtaining an alternative explanation for the apparent acceleration of the
cosmic expansion. This paved the way to the intriguing possibility that
unexplained aspects of the Cosmological Standard Model, such as dark energy,
dark matter or the inflation itself, could find a natural explanation within
fractal cosmology.

The data fit carried out in \cite{Cosmai:2018nvx} yielded to an evaluation $%
D=2.9\pm 0.02$, which is quite different from the previous result \cite%
{fract}. Such a gap might be due to the fact that the spatial extension of
the fractal structure is actually limited, since self-similarity disappears
beyond a certain \textquotedblleft greatness\textquotedblright\ scale. It is
conceivable that the hierarchy of super-super-clusters and void bubbles
comes to an end at some scale $L_{EG}$, suggestively named \textquotedblleft
End of Greatness\textquotedblright . Beyond such a scale length, the smaller
fractal figures appear just as juxtaposed, so that the mass law approximates
$M(r)\propto r^{3}$. An evaluation based on data fit has allowed to estimate
$L_{EG}\cong 100$ Mpc \cite{eg}.

Therefore, the total homogenized metric is described as a LTB metric below $%
L_{EG}$, and as a FLRW one beyond : this is the so-called \textit{Swiss
cheese model}, investigated by both \cite{Ribeiro:2008rs} and \cite%
{Cosmai:2018nvx} : the spherical fractal region around the center resembles
a ``Swiss cheese hole'', surrounded by an otherwise homogeneous distribution
of ``cheese''. The matching between LTB and FLRW metrics at $L_{EG}$ will be
dealt with the so-called \textit{Darmois junction} \cite{Darmois}.

\subsection{Perturbative cosmology and retarded potentials}

An apparently unrelated approach has been investigated in recent years by a
number of papers, such as \cite{sergio}, \cite{re} and \cite{re2}, and it
exploits retarded gravitational potentials, generated by matter
concentrations and resulting in distortions of the space-time metric. Such
distortions are genuine effects of Einstein's General Relativity, and they
can be interpreted as dark matter and/or dark energy phenomena, whose they
may provide an explanation, at least to a certain extent.

Necessarily, within this framework one must deal with an inhomogeneous and
anisotropic distribution of matter, such that the gravitational influence of
far objects does not cancel out due to Birkhoff Theorem \cite{Birkhoff}, as
it instead holds for the LTB model. Even if such metric distortions decrease
with the distance, one should appreciate a magnification effect due to the
very expansion of the Universe: the farther the source, the more it is in
the past, and higher the source $\tilde{\rho}$ itself. Such a magnification
has been confirmed by explicit computations e.g. in \cite{re} and in \cite%
{re2}, whose results highlighted a non-negligible effect on the metric,
despite the relative smallness of the matter inhomogeneity.

For the time being, an exact solution to the Einstein Equations with a very
asymmetrical source is still out of reach, and the search for retarded
potentials' solutions must necessarily rely on perturbation theory. Assuming
that the Cosmological Principle is a reasonably good approximation at the
visible Universe scale, the matter inhomogeneities are considered as small
ones,%
\begin{equation}
\tilde{\rho}(\underline{x};t):=\rho (\underline{x};t)-\bar{\rho}(t),
\end{equation}%
with respect to a homogeneous background $\bar{\rho}(t)$. Thus, the
inhomogeneities can be treated as a first order perturbation, with the
zeroth order approximation returning the Friedmann cosmology : the retarded
potentials arise as solutions to the linearization of Einstein Equations
around a FLRW background metric.

It can then be appreciated that the details concerning the spacial matter
distribution $\tilde{\rho}(\underline{x})$ are not so relevant, since the
crucial quantity is its average amount $\langle \tilde{\rho}\rangle $, which
provides a quantitative estimate of the \textquotedblleft total matter
inhomogeneity\textquotedblright\ \cite{re,re2}. In fact, this approach is
concerned with the total amount of dark matter and dark energy in the
Universe, not with their local distribution. As a consequence, the results
obtained in \cite{re,re2} are insensitive to the fractal nature of the
matter distribution, or to any other geometrical property matter can enjoy.
However, we should remark that a FLRW zeroth order background is just the
simplest starting point, but is quite inaccurate. A further investigation
should choose a less trivial background, such that mimics better the shape
of matter inhomogeneities.

\subsection{Retarded potentials and the fractal}

A tantalizing possibility is that deeper insights on cosmological matter
inhomogeneities may be gained by merging the two approaches presented in the
previous Sections. In the present work, we will indeed consider a fractal
matter distribution, \textit{at least} up to the $L_{EG}$ scale. We will
describe the resulting metric, and consider the ``Swiss cheese
homogenization''~as the zeroth order approximation, and we will then deal
with a first order perturbative description of the real fractal $\rho (%
\underline{x})$. Within the choice of a LTB background, the effects due to
retarded potentials will also be effectively dealt with, thus allowing for a
more reliable evaluation of cosmological parameters, such as the
cosmological constant and the dark matter amount.

Recent computations with retarded potentials have improved the explanation
of dark matter as well as of dark energy \cite{re2}. However, after \cite%
{Cosmai:2018nvx} it is known that a suitable LTB background can explain the
appearance of dark energy. It is thus reasonable to expect that a
combination of the above two approaches may result in considerable advances
in the explanation, at least to some non-negligible extent, of both dark
matter and dark energy.

Further improvements may also be expected to shed some light on the choice
of the homogeneous density $\bar{\rho}$, which turned out to be a tricky
feature within the perturbative approach based on retarded potentials. If $%
\bar{\rho}$ is taken as the average of $\rho $, it returns $\langle \tilde{%
\rho}\rangle =0$, which means no effects at all from a first order
calculation. On the other hand, in \cite{re2} it was chosen%
\begin{equation}
\bar{\rho}:=\min_{\underline{x}}\rho (\underline{x}),
\end{equation}%
or%
\begin{equation}
\bar{\rho}:=\max_{\underline{x}}\rho (\underline{x}),
\end{equation}%
but it is not yet clear if these are physically sensible choices. This issue
does not arise at all in the fractal approach, because no such a thing as
spacial averaging exists for a fractal, which is endowed with a lower and
lower average as the space region under consideration widens up; eventually,
the average tends to zero because of the void bubbles. Moreover, the growth
of void bubbles prevents the determination of a unique real fractal density $%
\Phi $. In fact, if one tries to define it from $M(r):=\Phi (r)r^{D}$, this
will result in $\Phi (r)$ oscillating undefinitely. Within a fractal, such
an issue can be overcome by choosing the minimum $\Phi $ of the oscillations
as the reference for the definition of the homogenization $\bar{\rho}(r)$.
One can appreciate that this procedure is physically meaningful, because the
perturbation $\tilde{\rho}(\underline{x})$ may have negative and positive
values here and there, but its average will certainly be positive, so that
first order effects will not vanish.

All in all, in this paper we aim at consistently determining the parameters
of our model: $D,\Phi ,L_{EG},\tilde{\rho}_{0},\rho _{\Lambda 0},\rho _{R0}$%
. They will be obtained by fitting the experimental data, such as the dark
matter effects and the luminosity distances of SNe Ia. Furthermore, local
metric distortions due to retarded effects will be compared to the expected
dark matter inside the single galaxy or cluster, thus discerning to what
extent they can effectively be explained as relativistic effects.

\subsection{\label{three epochs}The origin of the fractal, and the three
epochs}

The perturbative approach should concern the LTB approximation for ancient
times, when most perturbations were generated. The validity of the LTB model
over time would actually be an interesting issue \textit{per se}, because
the solution used so far is valid just around the current instant. Back in
time, for ancient times, we know that radiation dominates, and the evolution
of the metric gets distorted.

For what concerns the origin of the fractal distribution of matter, we put
forward the conjecture that it arises out as a consequence of the
matter-antimatter (M-AM) recombination process. In fact, as a tiny fraction
of matter survives the annihilation, it is conceivable that it was not
homogeneously distributed, but rather it is scattered only across those
regions in which the matter itself turned out to have a slightly larger
density. Before the recombination, the inhomogeneity of matter would be
mainly due to quantum uncertainty, being very small. However, after
recombination only a $\sim 10^{-9}$ fraction of the pre-existing matter
survives, and thus its inhomogeneity is magnified of a factor $\sim 10^{9}$.
For our purposes, we can suppose that matter was already distributed as a
fractal in very ancient times; in fact, this solves also the problem of
structure formation: dark matter is not actually needed, if matter was
sufficiently concentrated at the very beginning.

In our model, the evolution of the Universe is characterized in terms of
three different epochs, as follows.

\begin{enumerate}
\item \textit{Before M-AM recombination}. The Universe is well described by
FLRW, and quantum uncertainty is the unique source of perturbations.

\item \textit{M-AM recombination}. It generates a matter remnant with
fractal distribution, exhibiting a non-negligible inhomogeneity. It
generates a large amount of homogeneous radiation, as well.

\item \textit{After M-AM recombination}. At zeroth order, it is approximated
by a LTB Universe, starting with the dominance of a homogeneous radiation,
progressively fading away into an epoch in which fractal matter gets
dominant. The first order perturbations better approximate the actual
fractal, and they give rise to retarded distortions. The superposition of
these latter for all times effectively results into dark matter phenomena,
both globally and locally, as the fractal geometry causes a distortion of
the luminosity distances which appears as a Universe acceleration.
\end{enumerate}

\subsection{The \textquotedblleft Swiss cheese\textquotedblright\ metric}

In this paper we will consider a Swiss cheese metric:%
\begin{equation}
d\bar{s}^{2}=\left\{
\begin{array}{ll}
-dt^{2}+\frac{A^{\prime 2}}{f^2}dr^{2}+A(r;t)^{2}d\Omega ^{2}, & L_{G}\leq
r\leq L_{EG}; \\
&  \\
-dt^{2}+a(t)^{2}[dx^{2}+x^{2}d\Omega ^{2}], & x\geq L_{EG},%
\end{array}%
\right.  \label{swiss}
\end{equation}%
where the coordinate arbitrariness is fixed as
\begin{equation}
A(r;0)\equiv A_{0}(r):=r,\quad a(0)\equiv a_{0}:=1,  \label{fix}
\end{equation}%
and we use a prime ``$\ ^{\prime }\ $'' for $r$-derivative and a dot ``$\
\dot{}\ $'' for $t$-derivative.\newline
Today, the matter inside dominates and is homogenized as%
\begin{equation}
\bar{M}_{0}(r)|_{[L_{G};L_{EG}]}=\Phi r^{D}\qquad\Rightarrow\qquad \bar{\rho}%
_{0}(r)|_{[L_{G};L_{EG}]}=\frac{D}{4\pi }\Phi r^{D-3}.
\end{equation}%
The matter outside is already homogeneous, with some value%
\begin{equation}
\bar{\rho}_{0}(x)_{[L_{EG};\infty )}\equiv \bar{\rho}_{00}.
\end{equation}%
The fractal dimension $D$ can be measured as in \cite{fract}. It does not
deform the luminosity distances everywhere, but just until $L_{EG}$, which
can be coherent to the different measure in \cite{Cosmai:2018nvx}.

For $r<L_{G}$, the exact metric depends on the distribution of matter in a
galaxy. A first simplification is to consider the fractal of matter as made
of balls whose minimum radius is $L_{G}$; hence, the galaxy would be
approximated as a homogeneous sphere, and thus below $L_{G}$ another
Friedmann metric would arise.

We consider as $A(r;t)$ is also a FLRW metric during epoch 1, whence it
gains an inhomogeneity during epoch 2, and the epoch 3 sees the evolution of
fractal. From now on, we will try to describe such $A(r;t)$, especially
during epoch 3.

\section{Inadequacy of LTB during epoch 3}

\subsection{Pure matter}

A universe filled with only matter regulates the Friedman Equation outside as%
\begin{equation}
\left( \frac{\dot{a}}{a}\right) ^{2}=\frac{8}{3}\pi G\bar{\rho}_{00}a^{-3}.
\end{equation}

There are no singularities of density, thus the metric must be almost
everywhere twice derivable: $\bar{g}_{\mu \nu }\in C^{1}$. Such a
requirement contains the Darmois junction, which defines the dependence $%
x(r;t)$. These have especially the consequences%
\begin{align}  \label{darm}
&\bar{g}_{tr}\in C^1(L_{EG})\quad\ \Rightarrow \quad\ \dot{x}%
(L_{EG};t)\equiv0 \quad\ \Rightarrow \quad\ x(L_{EG};t)\equiv L_{EG}; \cr &
\bar{g}_{\Omega\Omega}\in C^0(L_{EG}) \quad\ \Rightarrow \quad\
A(L_{EG};t)=a(t)x(L_{EG};t)=L_{EG}a(t).
\end{align}

Within the fractal assumption $\bar{M}_{0}(r):=\Phi r^{D}$, and following
\cite{Ribeiro:2008rs} and \cite{Cosmai:2018nvx} setting $f:=1$, we get the
functions%
\begin{equation}
A(r;t)=r\left[ 1+\frac{3}{2}H_{0}(r)t\right] ^{\frac{2}{3}}=r\left[ 1+\frac{3%
}{2}\sqrt{2G\Phi }r^{\frac{D-3}{2}}t\right] ^{\frac{2}{3}}.  \label{A}
\end{equation}%
For Darmois (\ref{darm}), it yields to%
\begin{equation}
a(t)=\frac{1}{L_{EG}}A(L_{EG};t)=\left[ 1+\frac{3}{2}\sqrt{2G\Phi }L_{EG}^{%
\frac{D-3}{2}}t\right] ^{\frac{2}{3}}.
\end{equation}%
Moreover, by differentiating Darmois, one obtains%
\begin{gather}
h_{0}:=\frac{\dot{a}_{0}}{a_{0}}=\frac{\dot{A}_{0}(L_{EG})}{A_{0}(L_{EG})}%
=H_{0}(L_{EG})=\sqrt{2G\Phi }L_{EG}^{\frac{D-3}{2}}; \\
\Downarrow  \notag \\
a(t)=\left[ 1+\frac{3}{2}h_{0}t\right] ^{\frac{2}{3}},\quad
H_{0}(r)=h_{0}\left( \frac{r}{L_{EG}}\right) ^{\frac{D-3}{2}}.  \label{H0}
\end{gather}%
Moreover, by imposing the Friedmann equation to hold outside, the following
results are achieved :%
\begin{gather}  \label{F out}
\left( \frac{\dot{a}}{a}\right) ^{2}=\frac{8}{3}\pi G\bar{\rho}%
_{00}a^{-3}=h_0^2a^{-3}\qquad\Rightarrow\qquad \dot{a}^{2}=h_{0}^{2}a^{-1}%
\quad \text{s.t.}\quad h_{0}^{2}=\frac{8}{3}\pi G\bar{\rho}_{00}; \\
\Downarrow  \notag \\
a(t)=\left[ \frac{3}{2}h_{0}(t-t_{I})\right] ^{\frac{2}{3}}\qquad
\Rightarrow\qquad \left[ 1+\frac{3}{2}h_{0}t\right] ^{\frac{2}{3}}=\left[
\frac{3}{2}h_{0}(t-t_{I})\right] ^{\frac{2}{3}}; \\
\Downarrow  \notag \\
t_{I}=-\frac{2}{3}h_{0}^{-1},
\end{gather}%
and%
\begin{gather}
2G\Phi L_{EG}^{D-3}=H_{0}(L_{EG})^{2}=\frac{8}{3}\pi G\bar{\rho}_{00}; \\
\Downarrow  \notag \\
\bar{M}_{0}(L_{EG})=\Phi L_{EG}^{D}=\frac{4}{3}\pi L_{EG}^{3}\bar{\rho}_{00}.
\end{gather}%
Thus, the Swiss cheese metric has a time singularity at%
\begin{equation}
t_{S}=-\frac{2}{3}\left( \frac{L_{G}}{L_{EG}}\right) ^{\frac{3-D}{2}%
}h_{0}^{-1}>t_{I},  \label{t_s}
\end{equation}%
at which $A^{\prime }(L_{G};t_{S})$ goes to infinity. Here the validity of
our pure matter model reaches an end.\bigskip

\textbf{Remark 1.} Usually, the Big Bang is set at the time singularity of
the metric. However, for the pure matter model such a singularity depends on
$r$ :
\begin{equation}
t_{BB}(r)=-\frac{2}{3}\left( \frac{r}{L_{EG}}\right) ^{\frac{3-D}{2}%
}h_{0}^{-1},
\end{equation}%
such that $t_{S}$ (\ref{t_s}) is just the first instant without singularity
: $t_{S}:=\max_{r}t_{BB}(r)$. This result makes no sense, since the Big Bang
should be the same for all the Universe. Therefore, the pure matter does not
provide a satisfactory description, and multi-component model is needed. In
particular, a component with a larger $w$, such as radiation, will do the
job : if it dominates in the early Universe, with an initial homogeneous
density, it would grant the synchronicity of Big Bang for all $r$. This
reasoning implies that the pure matter Swiss cheese metric (\ref{swiss})
with (\ref{A}) can be a good approximation only near the current instant,
but generally the evolution must concern a multi-component model.

\subsection{The \textit{flat} LTB model}

A consistent description of the expansion of the Universe would involve many
components - namely matter, radiation, and eventually dark energy - and
their evolutions.

To this aim, we need to make $\bar{\rho}_{M}(r;t)$ explicit; the functional
dependence on time is obtained from \cite{Ribeiro:2008rs} to be%
\begin{gather}
8\pi G\bar{\rho}_{M}=\frac{F^{\prime }}{2A^{\prime 2}},\quad \text{s.t.}%
\quad F=2A\dot{A}^{2}; \\
\Downarrow  \notag \\
\bar{\rho}_{M}(r;t)=\frac{D}{\pi }\Phi r^{D-3}\frac{1}{%
[2+3H_{0}(r)t][2+DH_{0}(r)t]}.  \label{pre-still}
\end{gather}%
It should be remarked that $\bar{\rho}_{M}(r;t)$ goes as the inverse of the
volume ($\bar{\rho}_{M0}(r):=\bar{\rho}_{M}(r;0)$):%
\begin{equation}
\bar{\rho}_{M}(r;t)=\frac{4}{[2+3H_{0}(r)t][2+DH_{0}(r)t]}\bar{\rho}%
_{M}(r;0)=\frac{r^{2}}{A^{2}\left( r;t\right) A^{\prime }\left( r;t\right) }%
\bar{\rho}_{M0}(r),  \label{still}
\end{equation}%
as expected, since matter is still. On the other hand, the dark energy does
not depend on $t$, so its density reads%
\begin{equation}
\bar{\rho}_{\Lambda }(r;t)=\bar{\rho}_{\Lambda 0}\left( r\right) .
\label{dark}
\end{equation}%
In case it is a cosmological constant, it should also be independent of $r$.

Analogously to the FLRW model, one would expect that the radiation density
goes as%
\begin{equation}
\bar{\rho}_{R}\left( r;t\right) \propto \left( \frac{r^{2}}{A^{2}\left(
r;t\right) A^{\prime }\left( r;t\right) }\right) ^{\frac{4}{3}},
\label{conj}
\end{equation}%
but this should better be confirmed by a more detailed computation (cfr. (%
\ref{ress}) further below).

We will henceforth carry out a detailed treatment of the flat LTB model. The
LTB metric returns a diagonal Einstein tensor, with%
\begin{align}
G_{t}^{t} &=-\frac{\dot{A}}{A}\left( 2\frac{\dot{A}^{\prime }}{A^{\prime }}+%
\frac{\dot{A}}{A}\right) ; \\
G_{r}^{r} &=-2\frac{\ddot{A}}{A}-\frac{\dot{A}^{2}}{A^{2}}; \\
G_{\theta }^{\theta } &=G_{\varphi }^{\varphi }=-\frac{\ddot{A}^{\prime }}{%
A^{\prime }}-\frac{\ddot{A}}{A}-\frac{\dot{A}^{\prime }\dot{A}}{A^{\prime }A}%
.
\end{align}
Hence, also $T_{\mu \nu }$ is diagonal, implying still matter. Within the
assumption of mostly-plus signature and the symmetries of our system, the
energy-momentum tensor of a perfect fluid reads%
\begin{gather}
T_{\mu \nu }:=(\rho +p)U_{\mu }U_{\nu }+pg_{\mu \nu },~\text{with~}U_{\mu
}=\delta _{t\mu }; \\
\Downarrow  \notag \\
T_{t}^{t}=-(\rho +p)+p=-\rho ,\quad T_{r}^{r}=p,\quad T_{\theta }^{\theta
}=T_{\varphi }^{\varphi }=p.  \label{T}
\end{gather}%
Thus, three independent Einstein equations are obtained, namely :%
\begin{equation}
\begin{cases}
-\frac{\dot{A}}{A}\left( 2\frac{\dot{A}^{\prime }}{A^{\prime }}+\frac{\dot{A}%
}{A}\right) =-8\pi G\rho ; \\
\\
-2\frac{\ddot{A}}{A}-\frac{\dot{A}^{2}}{A^{2}}=8\pi Gp; \\
\\
-\frac{\ddot{A}^{\prime }}{A^{\prime }}-\frac{\ddot{A}}{A}-\frac{\dot{A}%
^{\prime }\dot{A}}{A^{\prime }A}=8\pi Gp.%
\end{cases}
\label{ein sys}
\end{equation}

\subsubsection{\label{Riccati}The Ricci equation as a Riccati equation, and
its solutions}

With a barotropic equation of state $\rho =\rho (p)$, one has four equations
for the three unknowns $A,\rho ,p$. This should imply some constraint on the
form of $A,\rho ,p$. Such a constraint can be obtained from the second and
third Einstein equations, as follows:
\begin{gather}
2\frac{\ddot{A}}{A}+\frac{\dot{A}^{2}}{A^{2}}=8\pi Gp=\frac{\ddot{A}^{\prime
}}{A^{\prime }}+\frac{\ddot{A}}{A}+\frac{\dot{A}^{\prime }\dot{A}}{A^{\prime
}A}; \\
\Downarrow  \notag \\
\frac{\ddot{A}}{A}+\frac{\dot{A}^{2}}{A^{2}}=\frac{\ddot{A}^{\prime }}{%
A^{\prime }}+\frac{\dot{A}^{\prime }\dot{A}}{A^{\prime }A}.  \label{PDE}
\end{gather}%
We can try to solve this non-linear PDE in $A$, which we will name \textit{%
Ricci equation}, and search for a set of self-consistent solutions.
Exploiting the definition%
\begin{equation}
H:=\frac{\dot{A}}{A},  \label{H}
\end{equation}

the identity (\ref{PDE}) can be rewritten in a very simple way,%
\begin{equation}
\dot{H}^{\prime }+3HH^{\prime }=0.  \label{H PDE}
\end{equation}%
For a general Universe, (\ref{H PDE}) constrains the possible matter,
radiation and/or dark energy content. It is easy to check that the solution
found in \cite{Cosmai:2018nvx} satisfies this PDE. Nevertheless, there is no
uniqueness proven for the solutions of (\ref{H PDE}), so we can search for
other, different solutions.

Now, (\ref{H PDE}) can be rewritten as
\begin{gather}
0=\partial _{r}\left( \dot{H}+\frac{3}{2}H^{2}\right) ; \\
\Downarrow  \notag \\
\dot{H}+\frac{3}{2}H^{2}=c(t),  \label{H PDE 2}
\end{gather}%
where the integration constant $c(t)$ does not depend on $r$. Eq. (\ref{H
PDE 2}) can be recognized to be a Riccati Equation. For $c(t)\equiv 0$, we
find again the solution in \cite{Cosmai:2018nvx}, namely%
\begin{equation}
H(r;t)=\frac{2H_{0}(r)}{2+3H_{0}(r)t}.  \label{c=0 sol}
\end{equation}%
But e.g. for a non-zero, constant $c(t)\equiv c$ we can find different
solutions. Calling $c:=\frac{3}{2}\tau ^{-2}$, we get
\begin{align}
H(r;t)=\frac 1\tau \tanh \frac {3t}{2\tau} +H_0(r),
\label{c non-zero constant sol}
\end{align}
while for negative $c=-\frac 32 \alpha^2$ one has the (quite unphysical)
solution
\begin{align}
H(r;t)=-\alpha \tan \frac {3\alpha t}{2} +H_0(r).
\end{align}

On the other hand, we observe that the second Einstein Eq. from (\ref{ein
sys}) depends only on $H$, thus exploiting (\ref{H PDE 2}) we can obtain the
following expression for th pressure $p$:%
\begin{equation}
8\pi Gp=-2\frac{\ddot{A}}{A}-\frac{\dot{A}^{2}}{A^{2}}=-2\dot{H}%
-3H^{2}=-2\left( \dot{H}+\frac{3}{2}H^{2}\right) =-2c(t).  \label{pre-p}
\end{equation}%
Therefore, the integration constant gets related to $p$ itself: $c(t)=-4\pi
Gp$, which implies that the total pressure \textit{must be homogeneous} at
any time:%
\begin{equation}
p(r;t)=p(t).  \label{p}
\end{equation}

\subsubsection{Conservation of four-momentum and separability of $w$'s}

Let us now study the conservation of the four-momentum. One can compute the
conservation of energy%
\begin{gather}
\dot{\rho}=-\left( \frac{\dot{A}^{\prime }}{A^{\prime }}+2\frac{\dot{A}}{A}%
\right) (\rho +p),  \label{energy cons}
\end{gather}%
and the conservation of momentum, which turn out to be%
\begin{gather}
p^{\prime }=0.  \label{momentum cons}
\end{gather}%
This equation is just a confirmation of the result (\ref{p}), expressing the
homogeneity of pressure, as it must be for a perfect fluid in a LTB \textit{%
flat} Universe.

Let us now consider a particular type of perfect fluid, namely a \textit{%
single-component} one, defined by $p:=w\rho $. The homogeneity of pressure
then immediately implies%
\begin{equation}
w\rho ^{\prime }=0.
\end{equation}%
We can thus conclude that we \textit{no single-component, inhomogeneous flat
LTB Universe can exist, unless such a component is matter}. It then turns
out that the two solutions for the single-component case in a flat LTB
Universe were actually already both studied: for $w=0$, the pure matter flat
LTB model, studied in \cite{Tolman}, \cite{Bonnor}, \cite{Ribeiro:2008rs}
and \cite{Cosmai:2018nvx}, is retrieved; for $\rho ^{\prime }=0$, one simply
obtained the well-known FLRW model.

The case of a \textit{multi-component} perfect fluid is more interesting. By
setting%
\begin{equation}
p=\sum_{w}p_{w}=\sum_{w}w\rho _{w},~\text{s.t.~}\rho =\sum_{w}\rho _{w},
\label{multi}
\end{equation}%
the conservation of momentum (\ref{momentum cons}) allows for
inhomogeneities to exist for any component $\rho _{w}$, but only if the
pressure inhomogeneities compensate each other,
\begin{equation}
\sum_{w}w\rho _{w}^{\prime }(r;t)=0.
\end{equation}%
On the other hand, the conservation of energy allows one to study each
component separately (i.e., by fixing the corresponding $w$); indeed, (\ref%
{energy cons}) and (\ref{multi}) yield%
\begin{equation}
\sum_{w}\dot{\rho}_{w}=-\left( \frac{\dot{A}^{\prime }}{A^{\prime }}+2\frac{%
\dot{A}}{A}\right) \sum_{w}(1+w)\rho _{w}.  \label{CoE}
\end{equation}%
Within the assumption of separation of components\footnote{%
We will see below that such an assumption would not hold during epoch 2
(cfr. Sec. \ref{epoch 2}).}, \textit{for each} component $w$ we find%
\begin{gather}
\partial _{t}\ln \rho _{w}=\frac{\dot{\rho}_{w}}{\rho _{w}}=-(1+w)\left(
\frac{\dot{A}^{\prime }}{A^{\prime }}+2\frac{\dot{A}}{A}\right)
=-(1+w)\partial _{t}(\ln A^{\prime }+2\ln A)=\partial _{t}\left(
A^{2}A^{\prime }\right) ^{-1-w},~\forall w; \\
\Downarrow  \notag \\
\rho _{w}(r;t)=\rho _{w0}(r)\left( \frac{A_{0}(r)^{2}A_{0}^{\prime }(r)}{%
A^{2}(r;t)A^{\prime }(r;t)}\right) ^{1+w},~~\forall w.  \label{ww}
\end{gather}%
For $w=0$ (matter), and choosing the radial coordinate s.t. $A_{0}(r)\equiv
r $ (cfr. (\ref{fix})), one retrieves (\ref{still}), namely%
\begin{equation}
\rho _{M}(r;t)=\rho _{M0}(r)\frac{r^{2}}{A^{2}(r;t)A^{\prime }(r;t)}.
\end{equation}%
For $w=1/3$ (radiation), Eq. (\ref{ww}) confirms the conjecture (\ref{conj}%
), namely :%
\begin{equation}
\rho _{R}(r;t)=\rho _{R0}(r)\left( \frac{r^{2}}{A^{2}\left( r;t\right)
A^{\prime }\left( r;t\right) }\right) ^{4/3}.  \label{ress}
\end{equation}%
For $w=-1$ (dark energy) the density is constant, and the previous result is
confirmed, namely $\rho _{\Lambda }(r;t)=\rho _{\Lambda 0}$.

To recap, in a flat LTB Universe with just matter and radiation, the
radiation must be homogeneous, and this holds also in presence of a
cosmological constant (i.e., of homogeneous dark energy). Note that, while
this has been assumed in previous papers (cfr. e.g. \cite{re}, \cite{re2}),
we here deduced it from the conservation of four-momentum.

\subsection{The \textquotedblleft approximation with epochs\textquotedblright%
}

No explicit, exact solutions are known for the Einstein field equations in
such a general case, with many components. Thus, we will resort to the
so-called \textquotedblleft approximation with epochs\textquotedblright\ :
at any $(r;t)$ we will consider as if there were just the dominating
component, neglecting the others.

We start by noticing that, even if radiation and dark energy are
homogeneous, the matter is not; therefore, it might well be that for some $r$
we could be in an epoch, whereas for some other $r$ we are already in
another one. We will consider the case of dominating matter further below
(after remark 2), and we will now focus on an evolution dominated by
radiation. Moreover, from now on we will not consider the dark energy
component in our calculations: they would be just more complicate, without
let a better understanding.

An homogeneous distribution of primordial radiation could be assumed, thus
giving rise to a Friedmannian expansion during the epoch dominated by
radiation:%
\begin{equation}
\dot{a}^{2}\cong h_{0}^{2}\Omega _{R0}a^{-2}\quad\ \Rightarrow\quad\ a(t)=%
\left[ 2\sqrt{\Omega _{R0}}h_{0}(t-t_{BB})\right] ^{1/2},\quad \text{s.t.}%
\quad \Omega _{R0}:=\frac{8\pi G}{3h_{0}^{2}}\rho _{R0},
\end{equation}%
with the radiation evolving as $\Omega _{R}(t)=\Omega _{R0}a^{-4}\left(
t\right) $.\bigskip

\textbf{Remark 2. }$\rho _{R0}:=\rho _{R}(r;t=0)|_{r\geq L_{EG}}$ is the
radiation density today \textit{beyond the End of Greatness }$L_{EG}$, at
which it is still uniform. Below $L_{EG}$, one can reasonably assume that $%
\rho _{R}(r;0)$ is not homogeneous, since it developed through an
inhomogeneous expansion. This would imply the current measurements of $%
\Omega _{R0}$ not to be reliable, since they would take place inside our
galaxy, and thus in a point of the cosmic fractal: these would be measures
of $\rho _{R}(L_{G};0)$, which could be quite different from the average
value $\rho _{R0}$. For instance, inside a void bubble, the density of the
cosmic background would undergo a completely different development.\bigskip

When $\Omega _{M}(t)\geq \Omega _{R}(t)$, one would switch to the epoch
dominated by matter. $\Omega _{M}$ must also depend on $r$ :%
\begin{equation}
\Omega _{M}(r;t):=\frac{8\pi G}{3h_{0}^{2}}\bar{\rho}_{M}(r;t),\quad \text{%
s.t.}\quad \bar{\rho}_{M}(r;t)\propto \left\{
\begin{array}{l}
a(t)^{-3},~t<t_{RM}(r); \\
\\
\frac{r^{2}}{A^{2}A^{\prime }},~t>t_{RM}(r).%
\end{array}%
\right.
\end{equation}%
For a fixed $r<L_{EG}$, the \textquotedblleft soldering
instant\textquotedblright\ $t_{RM}$ is defined as%
\begin{equation}
\Omega _{M}(r;t_{RM}):=\Omega _{R}(t_{RM})\quad\ \Leftrightarrow\quad\ \frac{%
D\Phi r^{D-3}}{\pi \lbrack 2+3H_{0}(r)t_{RM}][2+DH_{0}(r)t_{RM}]}\bar{\rho}%
_{M}(r;0)=\rho _{R}(r;t_{RM}).
\end{equation}

\subsubsection{\label{two epochs}\textquotedblleft Swiss
cheese\textquotedblright\ with two epochs}

Next, we will consider again the Swiss cheese metric, in order to describe a
radiation$+$matter Universe by soldering the corresponding two one-component
solutions together.

Let us consider first the \textit{outer} expansion, which is simpler. The
Friedmann Eq. (\ref{F out}) at $r>L_{EG}$ reads%
\begin{equation}
h^{2}:=\left( \frac{\dot{a}}{a}\right) ^{2}=h_{0}^{2}(\bar{\Omega}%
_{R00}a^{-4}+\bar{\Omega}_{M00}a^{-3}),\text{s.t.}\bar{\Omega}_{w00}:=\frac{%
\bar{\rho}_{w00}}{\bar{\rho}_{00}},\quad \bar{\rho}_{00}:=\frac{3h_{0}^{2}}{%
8\pi G}.
\end{equation}%
The outside matter density is related to the inside matter density $\bar{\rho%
}_{M0}(r)=\frac{D}{4\pi }\Phi r^{D-3}$ by%
\begin{equation}
\Phi =\frac{4}{3}\pi L_{EG}^{3-D}\bar{\rho}_{M00}.
\end{equation}%
If $\bar{\Omega}_{M00}>\bar{\Omega}_{R00}$, it holds that%
\begin{equation}
a(0):=1>a_{RM}:=\frac{\bar{\Omega}_{R00}}{\bar{\Omega}_{M00}}>a_{BB}:=0.
\end{equation}%
Thus, during both epochs, the evolution of the Universe can be approximated
as if there were only one component, i.e. the dominating one :%
\begin{equation}
a(t)=\left\{
\begin{array}{l}
\left( 2h_{0}(t-t_{BB})\right) ^{1/2},~t_{BB}\leq t\leq t_{RM}; \\
\\
\left( \frac{3}{2}h_{0}t+1\right) ^{2/3},~t_{RM}\leq t\leq 0.%
\end{array}%
\right.
\end{equation}%
Thus, one can compute the \textquotedblleft soldering
instant\textquotedblright\ $t_{RM}$ as follows:
\begin{equation}
\frac{3}{2}h_{0}t_{RM}+1=a_{RM}^{3/2}\Rightarrow t_{RM}=\frac{2}{3}%
h_{0}^{-1}[a_{RM}^{3/2}-1].  \label{sold-outside}
\end{equation}%
From the continuity of $a(t)$ at $t_{RM}$, one obtains also the homogeneity
for the Big Bang instant $t_{BB}(r)\equiv t_{BB}$.

Let us now consider the \textit{inner} expansion; for a fixed $r<L_{EG}$,
since the radiation epoch must be homogeneous, we know that the evolution is%
\begin{equation}
A(r;t)=\left\{
\begin{array}{l}
r\left( 2h_{0}(t-t_{BB})\right) ^{1/2},~t_{BB}\leq t\leq t_{RM}(r); \\
\\
r\left( \frac{3}{2}H_{0}(r)t+1\right) ^{2/3},~t_{RM}(r)\leq t\leq 0,%
\end{array}%
\right.
\end{equation}%
where we recalled the result (\ref{H0}). When we try to compute the
\textquotedblleft soldering instant\textquotedblright\ $t_{RM}(r)$ within
this regime, we can appreciate the inadequacy of the framework under
consideration in order to describe a Universe with radiation and matter;
indeed, we should impose the continuity of $A(r;t)$, and thus solve%
\begin{equation}
\left( 2h_{0}(t_{RM}(r)-t_{BB})\right) ^{3}=\left( \frac{3}{2}%
H_{0}(r)t_{RM}(r)+1\right) ^{4},  \label{IV deg}
\end{equation}%
which is a fourth degree algebraic equation. Surely, for $r\rightarrow
L_{EG}^{-}$, we will retrieve the expression of $t_{RM}$ (\ref{sold-outside}%
) computed within the outside expansion, because $H_{0}(L_{EG})=h_{0}$.
Nevertheless, let us consider the definition of $t_{RM}(r)$ as the instant
when the matter density and the radiation density are equal; by defining $x:=%
\frac{r}{L_{EG}} $, one can write%
\begin{gather}
\bar{\rho}_{R00}[2h_{0}(t-t_{BB})]^{-2}=\bar{\rho}_{R00}a^{-4}=\bar{\rho}%
_{R}(r;t)=\bar{\rho}_{M}(r;t)  \notag \\
=\bar{\rho}_{M0}(r)\frac{r^{2}}{A^{2}A^{\prime }}=\frac{D}{3}x^{D-3}\bar{\rho%
}_{M00}\left( 1+\frac{3}{2}x^{\frac{D-3}{2}}h_{0}t\right) ^{-1}\left( 1+%
\frac{D}{2}x^{\frac{D-3}{2}}h_{0}t\right) ^{-1} \\
\Downarrow  \notag \\
a_{RM}\left( 1+\frac{3}{2}x^{\frac{D-3}{2}}h_{0}t\right) \left( 1+\frac{D}{2}%
x^{\frac{D-3}{2}}h_{0}t\right) =\frac{D}{3}x^{D-3}[2h_{0}(t-t_{BB})]^{2}=%
\frac{D}{3}x^{D-3}\left( 1+\frac{3}{2}x^{\frac{D-3}{2}}h_{0}t\right) ^{8/3},
\label{resss}
\end{gather}%
where we used (\ref{IV deg}) in the last step of (\ref{resss}). As
mentioned, for $x\rightarrow 1^{-}$ one should find again $t=t_{RM}$, thus
obtaining%
\begin{gather}
\left( \frac{3}{D}a_{RM}\right) ^{3}\left( 1+\frac{D}{3}(a_{RM}^{3/2}-1)%
\right) ^{3}=(1+a_{RM}^{3/2}-1)^{5}; \\
\Downarrow  \notag \\
\left( \frac{3}{D}a_{RM}+a_{RM}(a_{RM}^{3/2}-1)\right)
^{3}=a_{RM}^{15/2}\Leftrightarrow \left( \frac{3}{D}-1\right) a_{RM}=0.
\end{gather}%
Thus, we obtain that only trivial solutions are allowed for consistency,
namely, the trivial FLRW solution $D=3$, or the pure matter solution $%
a_{RM}=0$.

\subsection{Inadequacy of the \textit{flat} LTB model}

We have found that the Swiss cheese metric, with inhomogeneous matter and
non-zero radiation, cannot be self-consistent when assuming a spatially flat
metric and still energy-matter. In other words, a spatially flat,
inhomogeneous LTB solution with still energy-matter must necessarily contain
only matter, and possibly some dark energy, whose evolution $\propto Vol^{0}$
allows to preserve the homogeneity (however, dark energy cannot dominate
near the Big Bang, which will necessarily be inhomogeneous in any such
model; cfr. remark 1 above).

By setting to zero the velocity field, the conservation of momentum implies
the homogeneity of pressure ($p^{\prime }=0$) at\textit{\ any} instant, so
that there are no forces. Within this framework, one can appreciate that the
inconsistency between inhomogeneous matter and non-zero radiation can be
traced back to the homogeneity of pressure. Indeed, since the matter has
vanishing pressure, the conservation of momentum yields homogeneous
radiation density, at any instant. But the expansion iself is inhomogeneous,
due to the matter inhomogeneity; as a consequence, even if the radiation is
homogeneous at a given instant, it will evolve inhomogeneously with the
expansion, thus breaking the conservation of momentum.

In a Universe undergoing a two-epochs evolution (as we are assuming in this
Section), the conservation of momentum approximately holds during both
epochs: as for the homogeneous expansion during the radiation-dominated
epoch, so for the zero pressure expansion during the matter-dominated epoch.
However, the \textquotedblleft two-epochs approximation" fails in proximity
of the \textquotedblleft soldering instant\textquotedblright\ $t_{RM}$,
namely when radiation and matter are about to be equal. In such an
intermediate period of time, the pressure is no more negligible, but the
expansion is still inhomogeneous. The inconsistency arises because the
conservation of momentum prevents the determination of a well-defined
\textquotedblleft soldering instant\textquotedblright\ $t_{RM}$.

It is here worth remarking that that this inconsistency cannot be solved by
adding other components, possibly aiming at compensating the inhomogeneity
of the pressure of radiation. Indeed, even if some other $\rho _{w}$ allows
to set $w\rho _{w}^{\prime }+\frac{1}{3}\rho _{R}^{\prime }=p^{\prime }=0$
for a given instant, this cannot hold for other instants, because the $w$
component evolves as $\propto (Vol)^{-1-w}$ with $w\neq \frac{1}{3}$,
whereas the radiation evolves $\propto Vol^{-4/3}$.

The above clashing of volumetric expansions implies that the consistent way
to add the radiation, or any other component with $w\neq 0,-1$, to the LTB
model, is \textit{at most} two-fold, as one could consider a non-vanishing
velocity field $v$ (yielding a fourth Einstein equation, the one sourced by
the component $T_{tr}$ of energy-momentum tensor), and/or a non-vanishing
spatial curvature.

\subsection{The \textit{non-flat} LTB model: $k=k(r;t)$ and $v\neq 0$}

Let us generalize the LTB metric by adding a non-vanishing spatial curvature
$k:=k(r)$,
\begin{equation}
ds^{2}=-dt^{2}+\frac{(A^{\prime })^{2}}{f^{2}}dr^{2}+A^{2}d\Omega ^{2},\quad
\text{s.t.}\quad f(r)^{2}=1-k(r)^{2}.  \label{LTB gen}
\end{equation}%
The treatment of this metric given in \cite{Tolman} yields the Einstein
tensor to be diagonal again; in particular, $G_{r}^{t}=0$. In turn, this
implies a diagonal energy-momentum tensor, and for a perfect fluid the
conservation of momentum yields the following result:%
\begin{equation}
0=\partial _{r}T_{r}^{r}+2\Gamma _{r\theta }^{\theta }(T_{r}^{r}-T_{\theta
}^{\theta })=p^{\prime }.
\end{equation}%
However, the\ aforementioned inconsistency plaguing the flat LTB Universe is
not (yet) resolved in such a non-flat Universe. In fact, a still
energy-matter evolves as $\propto (Vol)^{-1-w}$, with some dependence on $f$
in the formula of the volume $Vol$; consequently, the conservation of
momentum still allows only matter and dark energy within an inhomogeneous
Universe, still exhibiting an inhomogeneous Big Bang (cfr. remark 1 above).

Thus, one must necessarily consider a non-vanishing velocity field ($v\neq 0$%
) within a non-flat LTB Universe. Since (compare with \S 6.1)%
\begin{equation}
T_{r}^{t}=-(\rho +p)v\sqrt{1+\left( \frac{f}{A^{\prime }}v\right) ^{2}}\neq
0,
\end{equation}%
this would imply a non-vanishing $G_{r}^{t}$, again forbidden by \cite%
{Tolman}. The only way out is to consider a moving energy-matter ($v\neq 0$)
within a Universe with the most general type of (non-vanishing) spatial
curvature (although the spherical symmetry is required nevertheless), namely
$k=k\left( r;t\right) $, thus implying $f^{2}=1-k(r;t)^{2}=f(r;t)^{2}$.
Indeed, the $(tr)$-component of the Einstein Eqs. results to be%
\begin{equation}
\frac{A^{\prime }}{A}\frac{\dot{f}}{f}=-4\pi G(\rho +p)v\sqrt{1+\left( \frac{%
f}{A^{\prime }}v\right) ^{2}}.
\end{equation}%
Thus, we have four variables $A,\rho ,v,f$ for four Einstein Eqs. (namely,
the three diagonal components $(tt)$, $(rr)$, $(\theta \theta )$, and the
non-diagonal component $(tr)$).

\section{\label{epoch 2}Expansion during M-AM recombination}

\subsection{Inseparability of components}

Let us consider again (\ref{CoE}). In the treatment given above, we have
assumed the separation of (\ref{CoE}) into each of its $w$-components, and
we have obtained that an inhomogeneous LTB Universe with non-vanishing
radiation requires a non-zero velocity field and a spatial curvature $k$
depending both on $t$ and $r$. However, during the M-AM recombination
(corresponding to the epoch 2; cfr. Sec. \ref{three epochs}), the separation
of Eq. (\ref{CoE}) into its $w$-components is a sufficient \textit{but not
necessary} condition for the solution of (\ref{CoE}) itself. In general,
some mixing terms among the different components $w$'s can occur, as a
consequence of the recombination between matter and antimatter, in which a
huge quantity of $w=0$ (matter) component gets transformed into the $w=1/3$
(radiation) component\footnote{%
This may result into an overly simplified physical picture during M-AM
recombination, but nevertheless we search for a solution within this
framework.}.

For simplicity's sake, let us consider now the case with matter ($w=0$)
\textit{and} radiation ($w=1/3$) only\footnote{%
From now on, we will not put any dark energy as a component, since we want
to explain the luminosity distance observations as a consequence of the
fractal metric. It is possible an analogous model with a cosmological
constant, but it would further complicate the equations.}. Eq.s (\ref{multi}%
) and (\ref{momentum cons}) imply%
\begin{equation}
\rho _{R}=\rho _{R}(t)=3p(t).  \label{jjj}
\end{equation}%
Then, the equation of the conservation of the energy (\ref{CoE}) can be
written as%
\begin{gather}
\dot{\rho}_{M}+\dot{\rho}_{R}=-\left( \frac{\dot{A}^{\prime }}{A^{\prime }}+2%
\frac{\dot{A}}{A}\right) \left( \rho _{M}+\frac{4}{3}\rho _{R}\right) ; \\
\Updownarrow  \notag \\
\dot{\rho}_{M}=-\frac{\partial _{t}(A^{2}A^{\prime })}{A^{2}A^{\prime }}\rho
_{M}-\left[ 3\dot{p}+4\frac{\partial _{t}(A^{2}A^{\prime })}{A^{2}A^{\prime }%
}p\right] .  \label{energy}
\end{gather}%
It can be integrated as%
\begin{equation}
\rho _{M}(r;t)=\frac{\left[ \mathcal{K}_{M}(r)+\int_0^{t}\dot{p}(\tau
)A^{2}(r;\tau )A^{\prime }(r;\tau )d\tau \right] }{A^{2}(r;t)A^{\prime
}\left( r;t\right) }-4p(t),  \label{rho_M}
\end{equation}%
where $\mathcal{K}_{M}(r)=r^2[\rho_{M0}(r)+4p_0]$.

\subsection{Einstein equations (\textit{flat} LTB without dark energy)}

Having obtained the explicit functional dependence of the matter density and
its relation with the pressure, let us now try to solve the Einstein
equations (\ref{ein sys}) within the \textit{flat} LTB model\footnote{%
A detailed treatment of the Einstein equations for the \textit{non-flat} LTB
model with $k=k(r;t)$ and $v\neq 0$ will be given in Sec. \ref{Einstein gen}.%
}. By specifying only matter and radiation, and recalling (\ref{jjj}),
Einstein equations read%
\begin{equation}
\begin{cases}
\frac{\dot{A}^{2}}{A^{2}}+2\frac{\dot{A}^{\prime }\dot{A}}{A^{\prime }A}%
=8\pi G[\rho _{M}+3p(t)]; \\
\\
2\frac{\ddot{A}}{A}+\frac{\dot{A}^{2}}{A^{2}}=-8\pi Gp(t); \\
\\
\frac{\ddot{A}^{\prime }}{A^{\prime }}+\frac{\ddot{A}}{A}+\frac{\dot{A}%
^{\prime }\dot{A}}{A^{\prime }A}=-8\pi Gp(t),%
\end{cases}
\label{ein sys 2}
\end{equation}%
where we stressed the fact that the pressure depends only on time, as
expressed by (\ref{jjj}), which in turn guarantees the conservation of
momentum. From the treatment given in the previous Section, the conservation
of energy is given by(\ref{energy}), whereas the equation of state for
matter and radiation has been taken into account by specifying $\rho =\rho
_{M}+\rho _{R}=\rho _{M}+3p$.

From the treatment of Sec. \ref{Riccati}, we know that equating the second
and third Einstein equations, one obtains a Riccati equation (\ref{H PDE 2})
for the Hubble parameter $H$ (\ref{H}):%
\begin{equation}
\dot{H}+\frac{3}{2}H^{2}+4\pi Gp(t)=0,  \label{Riccati Eq}
\end{equation}%
where Eq. (\ref{pre-p}) has been recalled. We have discussed above the
solutions for $p\left( t\right) =0$ (vanishing pressure) and for $p(t)=p\neq
0$ (non-vanishing, constant pressure), respectively given by Eqs. (\ref{c=0
sol}) (obtained in \cite{Cosmai:2018nvx}) and (\ref{c non-zero constant sol}%
). Following the usual method to solve such a class of differential
equations (cfr. e.g. \cite{ric}), we define the auxiliary variable $y\left(
r;t\right) $ as follows:%
\begin{equation}
H=:\frac{2}{3}\frac{\dot{y}}{y},  \label{y}
\end{equation}%
in terms of which the Riccati equation (\ref{Riccati Eq}) becomes linear :%
\begin{equation}
\ddot{y}=-6\pi Gp(t)y.  \label{y eq}
\end{equation}%
It can be appreciated that $y$ provides an alternative description of the
expansion of Universe, in place of the coefficient $A\left( r;t\right) $;
indeed, by recalling (\ref{H}) and (\ref{y}), one gets $A^{3}=y^{2}$, and $%
A^{2}A^{\prime }\propto yy^{\prime }$.

Hence, one can rewrite the Eq. (\ref{energy}) of conservation of energy as%
\begin{equation}
\dot{\rho}_{M}=-\frac{\partial _{t}(yy^{\prime })}{yy^{\prime }}\rho _{M}-%
\left[ 3\dot{p}+4\frac{\partial _{t}(yy^{\prime })}{yy^{\prime }}p\right] .
\label{energy 2}
\end{equation}
Analogously, one can rewrite the other Einstein equations%
\begin{align}
&6\pi G[\rho _{M}+3p(t)]=\frac{\dot{y}^{\prime }\dot{y}}{y^{\prime }y}; \\
&-6\pi Gp(t)=\frac{\ddot{y}}{y}.
\end{align}%
By construction, the third Einstein equation from (\ref{ein sys 2}) is
equivalent to the second one via (\ref{y eq}): both of them are (\ref{y eq})
again. Thus, the Einstein system (\ref{ein sys 2}) can be rewritten in a
simpler way in terms of the $y$ function (\ref{y}) as follows:%
\begin{equation}
\left\{
\begin{array}{l}
\frac{\dot{y}^{\prime }\dot{y}}{y^{\prime }y}=6\pi G\left[ \rho _{M}+3p(t)%
\right] ; \\
\\
\ddot{y}=-6\pi Gp(t)y.%
\end{array}%
\right.  \label{ein sys 3}
\end{equation}
We observe that it is useless to substitute $\rho _{M}$ from the first
Einstein equation inside (\ref{energy}), since it gives again (\ref{y eq}).

Thus, we end up with the system (\ref{ein sys 3}) composed by two
independent PDE's in terms of the functions $y\left( r;t\right) $ (\ref{y})
and $\rho _{M}\left( r;t\right) $ (\ref{rho_M}), but the 1-variable function
$p(t)$ remains here undetermined. It is then evident that some other
condition is needed in order to obtain a consistent evolution of the
Universe; it is easy to realize that such a missing condition should be
provided by the law of transformation from matter to radiation as resulting
from the M-AM recombination, which we did not consider yet. {\color{red} FIN
QUI}

\subsection{New variables}

We observe that the linear Riccati equation (\ref{y eq}) does not actually
depend on $r$; thus, since it is a second order equation, its general
solution $y(r;t)$ will be given by a linear combination of two purely $t$%
-dependent functions $y_{1}(t)$ and $y_{2}(t)$, with $r$-dependent
coefficients,%
\begin{equation}
y(r;t):=c_{1}(r)y_{1}(t)+c_{2}(r)y_{2}(t),  \label{pre-yrt}
\end{equation}%
where%
\begin{equation}
\ddot{y}_{1,2}(t)=:-6\pi Gp(t)y_{1,2}(t).
\end{equation}%
The conditions at $t=0$ can be fixed e.g. by setting%
\begin{equation}
\begin{cases}
y_{1}(0)=0=\dot{y}_{2}(0); \\
\\
\dot{y}_{1}(0)=1=y_{2}(0),%
\end{cases}
\label{today}
\end{equation}%
which yields%
\begin{equation}
y(r;t)=A_{0}(r)^{3/2}\left[ \frac{3}{2}H_{0}(r)y_{1}(t)+y_{2}(t)\right] .
\label{yrt}
\end{equation}

Next, we notice the importance of the variable%
\begin{equation}
V:=y^{2}=A^{3}\Rightarrow yy^{\prime }=\frac{1}{2}V^{\prime },  \label{V}
\end{equation}%
which represents the volume of the sphere centred in $\vec{0}$ with radius $%
r $. By exploiting the definition (\ref{V}), the first Einstein equation of (%
\ref{ein sys 3}) can be recast in the following form (where $\rho =\rho
_{M}+3p$) :%
\begin{equation}
\dot{y}^{\prime }\dot{y}=3\pi G\rho V^{\prime },  \label{first Ein}
\end{equation}
whereas the equation of energy conservation (\ref{energy 2}) and the formula
of $\rho _{M}(r;t)$ (\ref{rho_M}) respectively acquire the following forms:%
\begin{equation}
\dot{\rho}=-\frac{\dot{V}^{\prime }}{V^{\prime }}\left( \rho +p\right) ,
\label{energy 3}
\end{equation}%
and%
\begin{gather}
\rho =\frac{1}{V^{\prime }}\left[\mathcal{K}_{M}(r)-\int^{t}p(\tau )\dot{V}%
^{\prime }\left( r;\tau \right) d\tau \right] ,
\end{gather}%
%
%
%
%

By inspecting Eq. (\ref{energy 3}), one can appreciate that an even better
variable to be used would be the total energy inside the sphere of radius $r$%
,%
\begin{eqnarray}
E(r;t) &:&=\int^{r}\rho (s;t)dV(s;t)=M(r;t)+3p(t)V(r;t),  \label{E}
\end{eqnarray}%
where%
\begin{equation}
M(r;t):=\int_{0}^{r}\rho _{M}(s;t)V(s;t)ds.  \label{M}
\end{equation}%
By virtue of the fact that definition (\ref{E}) implies%
\begin{equation}
E^{\prime }=\rho V^{\prime },  \label{E2}
\end{equation}%
the first Einstein equation (\ref{first Ein}) boils down to%
\begin{equation}
\dot{y}^{\prime }\dot{y}=3\pi GE^{\prime }.  \label{first Ein 2}
\end{equation}%
It can be integrated in $r$, obtaining%
\begin{equation}
\dot{y}^{2}=6\pi GE\Rightarrow \dot{E}=\frac{1}{3\pi G}\dot{y}\ddot{y}=-2y%
\dot{y}p=-p\dot{V},  \label{ven}
\end{equation}%
where the second Einstein equation (\ref{ein sys 3}) was used and definition
(\ref{V}) recalled. Finally, by integrating further in $t$, one obtains%
\begin{equation}
E(r;t)=E_{0}(r)-\int_{0}^{t}p(\tau )\dot{V}(r;\tau )d\tau ,  \label{EE}
\end{equation}%
where $E_{0}(r)=M_{0}(r)+3p_{0}$.

The evaluation of (\ref{ven}) today (i.e. for $t=0$) and the use of the time
derivative of (\ref{yrt}) yields the relation between $E_{0}(r)\equiv E(r;0)$
and $H_{0}(r)$ (by recalling the conditions (\ref{today})),%
\begin{equation}
\dot{y}^{2}\left( r;0\right) =6\pi GE_{0}\left( r\right) \Leftrightarrow
H_{0}^{2}(r)=\frac{8}{3}\pi G\frac{E_{0}\left( r\right) }{A_{0}(r)^{3}}=%
\frac{8}{3}\pi G\overline{\rho }_{0}(r),  \label{ven2}
\end{equation}%
where we defined $\overline{\rho }:=E/V=\bar{\rho}_{M}+3p$ as the average
density of energy inside the ball of radius $r$. The equation on the r.h.s.
of (\ref{ven2}) is a well known relation for FRW model, but in the framework
under consideration it depends on $r$. Analogously to FRW, we can define the
$\Omega $ parameters as follows :%
\begin{equation}
\Omega _{M}(r;t):=\frac{8\pi GM\bar{\rho}_{M}}{3H^{2}},\quad \Omega
_{R}(r;t):=\frac{8\pi Gp}{H^{2}}\quad \text{s.t.}\quad \Omega
_{M0}(r)+\Omega _{R0}(r):=1,  \label{Omegas}
\end{equation}%
where $\Omega _{M0}(r)\equiv \Omega _{M}(r;0)$ and $\Omega _{R0}(r)\equiv
\Omega _{R}(r;0)$.

\subsection{General form}

By plugging the time derivative of $y(r;t)$ (\ref{yrt}) into (\ref{ven}),
one obtains%
\begin{equation}
6\pi GE(r;t)=\dot{y}^{2}(r;t)=A_{0}(r)^{3}\left[ \frac{3}{2}H_{0}(r)\dot{y}%
_{1}(t)+\dot{y}_{2}(t)\right]
^{2}=A_{0}(r)^{3}[H_{0}(r)^{2}T_{2}(t)+H_{0}(r)T_{1}(t)+T_{0}(t)],
\end{equation}%
where%
\begin{equation}
T_{2}\left( t\right) :=\frac{9}{4}\dot{y}_{1}^{2}\left( t\right) ,\quad
T_{1}\left( t\right) :=3\dot{y}_{1}\left( t\right) \dot{y}_{2}\left(
t\right) ,\quad T_{0}\left( t\right) :=\dot{y}_{2}^{2}\left( t\right) .
\label{Ti}
\end{equation}%
On the other hand, by exploiting the first Einstein equation of (\ref{ein
sys 3}) an recalling the definition (\ref{V}), one could get a similar
expression for $\rho $, but it turns out to be non-polynomial,%
\begin{equation}
6\pi G\rho =\frac{\dot{y}^{\prime }\dot{y}}{y^{\prime }y}=\frac{\left[
3H_{0}(r)\dot{y}_{1}(t)+2\dot{y}_{2}(t)\right] \left[ 3H_{0}(r)A_{0}^{\prime
}(r)\dot{y}_{1}(t)+2A_{0}^{\prime }(r)\dot{y}_{2}(t)+2A_{0}(r)H_{0}^{\prime
}(r)\dot{y}_{1}(t)\right] }{\left[ 3H_{0}(r)y_{1}(t)+2y_{2}(t)\right] \left[
3H_{0}(r)A_{0}^{\prime }(r)y_{1}(t)+2A_{0}^{\prime
}(r)y_{2}(t)+2A_{0}(r)H_{0}^{\prime }(r)y_{1}(t)\right] }.
\end{equation}

\subsubsection{Pure matter}

Let us consider the pure matter case : $p\equiv 0$. From the second Einstein
equation of (\ref{ein sys 3}), one obtains%
\begin{equation}
y_{1}(t)=t,~~y_{2}=1  \label{y-matter}
\end{equation}%
yielding to%
\begin{equation}
A(r;t)=y^{2/3}(r;t)=A_{0}(r)\left[ \frac{3}{2}H_{0}(r)t+1\right] ^{2/3}.
\end{equation}%
Since $p=0$, Eqs. (\ref{E}) and (\ref{EE}) imply that%
\begin{equation}
\dot{M}=\dot{E}=0,
\end{equation}%
and the first Einstein equation of the system (\ref{ein sys 3}) simplifies
down to%
\begin{gather}
\rho _{M}\left( r;t\right) =\rho _{M0}\left( r\right) \frac{A_{0}^{\prime
}(r)}{\left[ A_{0}^{\prime }(r)\left( \frac{3}{2}H_{0}(r)t+1\right)
+A_{0}(r)H_{0}^{\prime }(r)t\right] \left[ \frac{3}{2}H_{0}(r)t+1\right] },
\end{gather}%
such that%
\begin{equation}
\rho _{M0}\left( r\right)=\frac{1}{4\pi G}\frac{H_{0}(r)\left[ \frac{3}{2}%
A_{0}^{\prime }(r)H_{0}(r)+A_{0}(r)H_{0}^{\prime }(r)\right] }{A_{0}^{\prime
}(r)}.
\end{equation}%
This is consistent with what we already know. The $T$ functions (\ref{Ti})
read
\begin{equation*}
T_{2}=\frac{9}{4},\quad T_{1}=0,\quad T_{0}=0.
\end{equation*}

\subsubsection{Pure radiation}

On the other hand, in the case of pure radiation $\rho _{M}\equiv 0$, Eqs. (%
\ref{E}) and (\ref{ven}) imply%
\begin{equation}
E=3pV\Rightarrow \dot{E}=3(\dot{p}V+p\dot{V})=-p\dot{V}\Leftrightarrow \frac{%
\dot{V}}{V}=-\frac{3}{4}\frac{\dot{p}}{p} \Leftrightarrow
V(r;t)=A_{0}(r)^{3}p(t)^{-3/4},
\end{equation}%
and by recalling (\ref{V}) one obtains%
\begin{equation}
A_{0}(r)^{3}p(t)^{-3/4}=V:=y^{2}=A_{0}(r)^{3}\left( \frac{3}{2}%
H_{0}(r)y_{1}(t)+y_{2}(t)\right) ^{2}.  \label{1}
\end{equation}%
And therefore, in this case it holds that%
\begin{eqnarray}
H_{0}(r) &\equiv &H_{0};  \label{3} \\
\frac{3}{2}H_{0}y_{1}(t)+y_{2}(t) &=&p(t)^{-3/8}.  \label{4}
\end{eqnarray}%
From%
\begin{equation}
H_{0}^{2}=\frac{8}{3}\pi G\frac{E_{0}\left( r\right) }{A_{0}(r)^{3}}=8\pi
Gp_{0}\Leftrightarrow p_{0}=\frac{H_{0}^{2}}{8\pi G},
\end{equation}%
one gets%
\begin{equation}
p\left( t\right) =\frac{1}{32\pi G}\left( t+\frac{1}{2H_{0}}\right) ^{-2}.
\end{equation}%
This result allows to explicitly solve the second Einstein equation of (\ref%
{ein sys 3}) yielding that%
\begin{gather}
y_{1}\left( t\right) =\frac{(2H_{0}t+1)^{3/4}-(2H_{0}t+1)^{1/4}}{H_{0}}%
,\quad y_{2}\left( t\right) =\frac{3(2H_{0}t+1)^{1/4}-(2H_{0}t+1)^{3/4}}{2},
\end{gather}%
finally leading to the following expression :%
\begin{eqnarray}
y &=&A_{0}(r)^{3/2}\left[ \frac{3}{2}H_{0}\frac{1}{H_{0}}\left(
(2H_{0}t+1)^{3/4}-(2H_{0}t+1)^{1/4}\right) +\frac{1}{2}\left(
3(2H_{0}t+1)^{1/4}-(2H_{0}t+1)^{3/4}\right) \right]  \notag \\
&=&A_{0}(r)^{3/2}(2H_{0}t+1)^{3/4},  \label{y-radiation}
\end{eqnarray}%
implying%
\begin{equation}
A\left( r;t\right) =y\left( r;t\right) ^{2/3}=A_{0}(r)\sqrt{2H_{0}t+1},
\end{equation}%
in which we recognize a feature of the FLRW model with pure radiation.

\subsubsection{Beyond pure models}

The two functional forms (\ref{y-matter}) and (\ref{y-radiation}),
respectively concerning the cases of pure matter and pure radiation can be
recognized to belong to a more general family of solutions\footnote{%
We should bear in mind that, physically, the \textquotedblleft right" $p(t)$
depends on the M-AM recombination law.}, of the form
\begin{equation}
y\propto (t+\theta )^{\frac{1\pm a}{2}},~~\text{s.t.~}\left\{
\begin{array}{l}
a=1~\text{for~pure~matter}; \\
\\
a=\frac{1}{2}~\text{for~pure~radiation}.%
\end{array}%
\right.
\end{equation}%
Again, the second Einstein equation of (\ref{ein sys 3}) implies%
\begin{gather}
p(t)=\frac{1-a^{2}}{24\pi G}(t+\theta )^{-2}.
\end{gather}%
By recalling the definition (\ref{Omegas}) of $\Omega _{R}(r;t)$, one then
obtains%
\begin{equation}
\frac{1-a^{2}}{24\pi G\theta ^{2}}=p(0)=\frac{\Omega _{R0}(r)H_{0}(r)^{2}}{%
8\pi G}\Leftrightarrow \theta ^{-1}\left( r;a\right) =H_{0}(r)\sqrt{\frac{3}{%
1-a^{2}}\Omega _{R0}(r)}.  \label{theta}
\end{equation}%
Note that such a result implies that in general $\theta $ does depend on $r$
(as well as on the parameter $a$).

We should now remember that we are considering the expansion of the Universe
during M-AM recombination only in presence of matter and radiation (namely,
we are disregarding the contribution of dark energy, for simplicity's sake).
Thus, $\rho _{M}\geq 0$ and $p\geq 0$ always, which imply $|a|\leq 1$.
Furthermore, it is reasonable to assume $\theta >0$, so that the Big Bang
happened in some past instant $t_{BB}=-\theta $. Within these assumptions,
the expression of $y(r;t)$ for the family of solutions under consideration
reads%
\begin{eqnarray}
y_{1}\left( t;a\right) &=&\frac{\theta }{\alpha }\left[ \left( \frac{t}{%
\theta }+1\right) ^{\frac{1+a}{2}}-\left( \frac{t}{\theta }+1\right) ^{\frac{%
1-a}{2}}\right] ; \\
y_{2}\left( t;a\right) &=&\frac{1}{2\alpha }\left[ (a-1)\left( \frac{t}{%
\theta }+1\right) ^{\frac{1+\alpha }{2}}+(a+1)\left( \frac{t}{\theta }%
+1\right) ^{\frac{1-\alpha }{2}}\right] ,
\end{eqnarray}%
which finally allows one to explicitly write down the functional form of the
$a$-parametrized family of solutions under consideration :%
\begin{equation}
y(r;t;a)=\frac{A_{0}^{3}(r)}{2a}\left[ \left( a+\sqrt{3\frac{1-a^{2}}{\Omega
_{R0}(r)}}-1\right) \left( \frac{t}{\theta }+1\right) ^{\frac{1+a}{2}%
}+\left( a-\sqrt{3\frac{1-a^{2}}{\Omega _{R0}(r)}}+1\right) \left( \frac{t}{%
\theta }+1\right) ^{\frac{1-a}{2}}\right] ,
\end{equation}%
where $\theta =\theta \left( r;a\right) $ given by (\ref{theta}). In turn,
this implies the formula%
\begin{equation}
A\left( r;t\right) =y^{2/3}(r;t)=\frac{A_{0}^{2}(r)}{\left( 2a\right) ^{2/3}}%
\left[ \left( a+\sqrt{3\frac{1-a^{2}}{\Omega _{R0}(r)}}-1\right) \left(
\frac{t}{\theta }+1\right) ^{\frac{1+a}{2}}+\left( a-\sqrt{3\frac{1-a^{2}}{%
\Omega _{R0}(r)}}+1\right) \left( \frac{t}{\theta }+1\right) ^{\frac{1-a}{2}}%
\right] ^{2/3}.
\end{equation}

\section{\label{Einstein gen}The Lema\^{\i}tre model}

Now we consider again the epoch 3, for which we saw the LTB solution is not
general enough. Thus, in this section we will use its generalization, called
the Lema\^{\i}tre model. It was described \textit{e.g.} in \cite%
{Bolejko:2011jc}. We choose the coordinates which diagonalize the metric
tensor, and we redefine $t$ with respect to eq. (7) of \cite{Bolejko:2011jc}
in order to get $g_{tt}:=-1$, which will mean that the energy-matter has
some radial velocity $U_{\mu}$. Hence, in our gauge the metric results to be%
\begin{equation}
ds^2=-dt^2+\left(\frac{A^{\prime }}{f}\right)^2dr^2+A^2(d\theta^2+\sin^2%
\theta d\phi^2)  \label{non-flat}
\end{equation}%
where the spatial curvature is $k(r;t)=\sqrt{1-f(r;t)^{2}}$, as (\ref{LTB
gen}).

\subsection{Einstein equations}

We will now adopt the tetrad formalism, in which $ds^{2}=\eta
_{ab}e^{a}\otimes e^{b}$, and which allows us to compute the \textit{Vielbein%
} as%
\begin{equation}
e^{0}=dt,\quad e^{1}=\frac{A^{\prime }}{f}dr,\quad e^{2}=Ad\theta ,\quad
e^{3}=A\sin \theta d\phi.
\end{equation}%
We can compute the Einstein tensor,%
\begin{equation}
\begin{array}{l}
G_{0}^{0}=-2\frac{\dot{A}^{\prime }\dot{A}}{A^{\prime }A}-\frac{\dot{A}^{2}}{%
A^{2}}-\frac{k^{2}}{A^{2}}+2\frac{f^{\prime }f}{A^{\prime }A}+2\left( \frac{%
\dot{A}}{A}-\frac{\dot{f}}{f}\right) \frac{\dot{f}}{f}; \\
\\
G_{0}^{1}=-2\frac{\dot{f}}{A}; \\
\\
G_{1}^{1}=-2\frac{\ddot{A}}{A}-\frac{\dot{A}^{2}}{A^{2}}-\frac{k^{2}}{A^{2}}%
+2\frac{\dot{f}^{2}}{f^{2}}; \\
\\
G_{2}^{2}=G_{3}^{3}=-\frac{\ddot{A}}{A}-\frac{\ddot{A}^{\prime }}{A^{\prime }%
}-\frac{\dot{A}^{\prime }\dot{A}}{A^{\prime }A}+\frac{f^{\prime }f}{%
A^{\prime }A}+\frac{\ddot{f}}{f}+\left( \frac{2\dot{A}^{\prime }}{A^{\prime }%
}+\frac{\dot{A}}{A}\right) \frac{\dot{f}}{f}.%
\end{array}%
\end{equation}

On the other hand, in presence of a non-vanishing velocity field, the
energy-momentum tensor reads%
\begin{equation}
T_{b}^{a}=(\rho +p)U^{a}U_{b}+p\delta _{b}^{a}\quad \text{s.t.}\quad U_{a}=%
\sqrt{v^{2}+1}e^{0}+ve^{1}=\sqrt{v^{2}+1}dt+v\frac{A^{\prime }}{f}dr,
\label{T1}
\end{equation}%
namely%
\begin{equation}
\begin{array}{l}
T_{0}^{0}=-(\rho +p)(v^{2}+1)+p=-\rho -v^{2}(\rho +p); \\
\\
T_{0}^{1}=v\sqrt{v^{2}+1}(\rho +p); \\
\\
T_{1}^{1}=p+v^{2}(\rho +p); \\
\\
T_{2}^{2}=T_{3}^{3}=p.%
\end{array}
\label{T2}
\end{equation}

Thus, we can finally write the Einstein equations for the Lema\^{\i}tre
model metric (\ref{non-flat}) with $k=k(r;t)$ and $v\neq 0$ :
\begin{equation}
\begin{cases}
2\frac{\dot{A}^{\prime }\dot{A}}{A^{\prime }A}+\frac{\dot{A}^{2}}{A^{2}}+%
\frac{k^{2}}{A^{2}}-2\frac{f^{\prime }f}{A^{\prime }A}-2\left( \frac{\dot{A}%
}{A}-\frac{\dot{f}}{f}\right) \frac{\dot{f}}{f}=8\pi G\rho +8\pi Gv^{2}(\rho
+p); \\
\\
\frac{\dot{f}}{A}=-4\pi Gv\sqrt{v^{2}+1}(\rho +p); \\
\\
2\frac{\ddot{A}}{A}+\frac{\dot{A}^{2}}{A^{2}}+\frac{k^{2}}{A^{2}}-2\frac{%
\dot{f}^{2}}{f^{2}}=-8\pi Gp-8\pi Gv^{2}(\rho +p); \\
\\
\frac{\ddot{A}}{A}+\frac{\ddot{A}^{\prime }}{A^{\prime }}+\frac{\dot{A}%
^{\prime }\dot{A}}{A^{\prime }A}-\frac{f^{\prime }f}{A^{\prime }A}-\frac{%
\ddot{f}}{f}-\left( \frac{2\dot{A}^{\prime }}{A^{\prime }}+\frac{\dot{A}}{A}%
\right) \frac{\dot{f}}{f}=-8\pi Gp.%
\end{cases}
\label{ein sys non-flat}
\end{equation}%
Notice that we don't express them in terms of the M variable, defined in eq.
(10) of \cite{Bolejko:2011jc}. We stress that M is not the \lq empirical
amount of mass\rq we defined in \S 3.3 and we will use again in \S 4.2.

The velocity $v$ represents the matter which falls on itself. We can assume
that it will be small almost always, w.r.t. $c=1$. Thus, we can approximate (%
\ref{ein sys non-flat}) up to the first order in $v$.%
\begin{equation}
\begin{cases}
2\frac{\dot{A}^{\prime }\dot{A}}{A^{\prime }A}+\frac{\dot{A}^{2}}{A^{2}}+%
\frac{k^{2}}{A^{2}}-2\frac{f^{\prime }f}{A^{\prime }A}-2\frac{\dot{A}}{A}%
\frac{\dot{f}}{f}=8\pi G\rho+o(v); \\
\\
\dot{f}=-4\pi GvA(\rho +p)+o(v); \\
\\
2\frac{\ddot{A}}{A}+\frac{\dot{A}^{2}}{A^{2}}+\frac{k^{2}}{A^{2}}=-8\pi
Gp+o(v); \\
\\
\frac{\ddot{A}}{A}+\frac{\ddot{A}^{\prime }}{A^{\prime }}+\frac{\dot{A}%
^{\prime }\dot{A}}{A^{\prime }A}-\frac{f^{\prime }f}{A^{\prime }A}-\frac{%
\ddot{f}}{f}-\left( \frac{2\dot{A}^{\prime }}{A^{\prime }}+\frac{\dot{A}}{A}%
\right) \frac{\dot{f}}{f}=-8\pi Gp;%
\end{cases}
\label{ein sys non-flat 3}
\end{equation}%
where we used than $\dot{f}=O(v)$ from the second equation. Moreover, the
first equation can be rewritten in the more compact form%
\begin{equation}
\partial_r[A(\dot{A}^2+k^2)]=8\pi G\frac{A^{\prime 2}}{f}[(f-\dot{A}v)\rho-%
\dot{A}vp].
\end{equation}

\subsection{Conservation laws}

In order to write the energy-momentum conservation, by recalling the
energy-momentum tensor (\ref{T1})-(\ref{T2}), one can approximate it as
follows:%
\begin{equation}
U_{\mu }=\sqrt{v^{2}+1}dt+v\frac{A^{\prime }}{f}dr\Rightarrow U^{\mu }=-%
\sqrt{v^{2}+1}\partial _{t}+v\frac{f}{A^{\prime }}\partial _{r},
\end{equation}%
which implies%
\begin{equation}
\begin{array}{l}
T_{t}^{t}=-\rho +o(v); \\
\\
T_{r}^{t}=-v\frac{A^{\prime }}{f}(\rho +p)+o(v); \\
\\
T_{t}^{r}=v\frac{f}{A^{\prime }}(\rho +p)+o(v); \\
\\
T_{r}^{r}=p+o(v); \\
\\
T_{\theta }^{\theta }=p=T_{\phi }^{\phi }.%
\end{array}%
\end{equation}

Hence, the conservation of energy reads%
\begin{gather}
\dot{\rho}=\left[ -\frac{\partial _{t}(A^{\prime}A^2/f)}{(A^{\prime}A^2/f)}+v%
\frac{f}{A^{\prime }}\frac{\partial _{r}(A^{\prime}A^2/f)}{(A^{\prime}A^2/f)}%
+\partial _{r}\left( v\frac{f}{A^{\prime }}\right) \right] (\rho +p)+v\frac{f%
}{A^{\prime }}(\rho ^{\prime }+p^{\prime })+o(v),
\end{gather}%
whereas the conservation of momentum is%
\begin{gather}
p^{\prime }=\left[ 2\frac{v}{A^{2}}\partial _{t}\left( \frac{A^{\prime}A^2}{f%
}\right) +\frac{A^{\prime}A^2}{f}\partial _{t}\left( \frac{v}{A^{2}}\right) %
\right] (\rho +p)+v\frac{A^{\prime }}{f}(\dot{\rho}+\dot{p})+o(v).
\label{cons-mom}
\end{gather}

We can rewrite the conservation laws by calling $\frac{A^{\prime}A^2}{f}%
:=V^{\prime }$, where $V(r;t)$ is the volume inside the sphere of radius $r$%
. The conservation of energy simplifies%
\begin{gather}
\partial _{t}(V^{\prime }\rho )=\partial _{r}\left[ A^{2}v(\rho +p)\right] -%
\dot{V}^{\prime }p.  \label{Et'}
\end{gather}%
The l.h.s. of (\ref{Et'}) is related to the total energy inside the sphere,
defined by (\ref{E}) as $E(r;t):=\int_{0}^{r}\rho (s;t)V^{\prime }(s;t)ds$,
such that (\ref{E2}) holds, namely $E^{\prime }=V^{\prime }\rho $. Within
the assumption of separation of (\ref{Et'}) into its $w$-components (which
holds with a good approximation after the M-AM recombination, when the
transformations occur only in the stars), one can write%
\begin{equation}
\dot{E}^{\prime }=\partial _{r}\left[ A^{2}v(\rho +p)\right] -\dot{V}%
^{\prime }p\Rightarrow \dot{E}_{w}^{\prime }=(1+w)\partial _{r}\left[ v\frac{%
f}{A^{\prime }}E_{w}^{\prime }\right] -w\frac{\dot{V}^{\prime }}{V^{\prime }}%
E_{w}^{\prime },~\forall w.
\end{equation}%
Indeed, for the static case $v=0$ we have just the volume deformation $%
E_{w}^{\prime }\propto (V^{\prime })^{-w}$. For the general case, the matter
component has a particularly simple law,%
\begin{equation}
\dot{M}^{\prime }=\partial _{r}\left[ v\frac{f}{A^{\prime }}M^{\prime }%
\right] \Rightarrow \dot{M}=v\frac{f}{A^{\prime }}M^{\prime }.
\end{equation}

Moreover, the conservation of momentum (\ref{cons-mom}) becomes%
\begin{gather}
V^{\prime }p^{\prime }=\partial _{t}\left[ v\frac{A^{\prime }}{f}V^{\prime
}(\rho +p)\right] .
\end{gather}

\subsection{General system}

The Lema\^{\i}tre model with $k=k(r;t)$ and $v\neq 0$, filled by matter and
radiation only, is described by five independent PDEs, which one can write
at the first order in $v$ as follows:%
\begin{equation}
\begin{cases}
\dot{f}=-4\pi GvA\left( \rho _{M}+\frac{4}{3}\rho _{R}\right) ; \\
\\
2\ddot{A}A+\dot{A}^{2}+1-f^{2}=-\frac{8}{3}\pi GA^{2}\rho _{R}; \\
\\
\partial _{t}(V^{\prime }\rho _{M})=\partial _{r}(v\frac{f}{A^{\prime }}%
\cdot V^{\prime }\rho _{M}); \\
\\
\partial _{t}[(V^{\prime 4/3}\rho _{R}]=\frac{4}{3}\sqrt[3]{V^{\prime }}%
\partial _{r}(v\frac{f}{A^{\prime }}\cdot V^{\prime }\rho _{R}); \\
\\
\rho _{R}^{\prime }=3\partial _{t}\left( v\frac{A^{\prime }}{f}\right) \rho
_{M}+4\sqrt[3]{V^{\prime }}\partial _{t}\left( \frac{vA^{\prime }}{f\sqrt[3]{%
V^{\prime }}}\right) \rho _{R}.%
\end{cases}
\label{sys}
\end{equation}%
The independent variables are $A,f,v,\rho _{M},\rho _{R}$, and the quantity $%
V:=\int_{0}^{r}\frac{A^{\prime}A^2}{f}$ has been defined. Moreover, the last
three conservation laws of (\ref{sys}) can alternatively be expressed as%
\begin{equation}
\begin{cases}
\dot{M}=v\frac{f}{A^{\prime }}M^{\prime }; \\
\\
\partial _{r}\left[ 4v\frac{f}{A^{\prime }}E_{R}^{\prime }-3\dot{E}_{R}%
\right] =\frac{\dot{V}^{\prime }}{V^{\prime }}E_{R}^{\prime }; \\
\\
E_{R}^{\prime \prime }=\left[ \frac{V^{\prime \prime }}{V^{\prime }}+4\sqrt[3%
]{V^{\prime }}\partial _{t}\left( \frac{vA^{\prime }}{f\sqrt[3]{V^{\prime }}}%
\right) \right] E_{R}^{\prime }+3\partial _{t}\left( v\frac{A^{\prime }}{f}%
\right) M^{\prime },%
\end{cases}%
\end{equation}%
where are defined the total quantities $E_{w}:=\int_{0}^{r}V^{\prime }\rho
_{w}$.

\subsection{Approximated models}

The PDE system in Sec. 4.3 is quite difficult to solve. The searched
solution fulfills a condition at the initial time $t_{R}$ ($A=ra(t)$, $%
1-f^{2}=Kr^{2}$, $\rho _{R}=\rho _{R}(t)$) and some other constraints at the
final instant ($A=r$, $f=1$, $M=\Phi r^{D}$). Hence, we cannot even exploit
a numerical approach, at least at the first step, because it would require a
complete set of conditions at a certain instant, e.g. the initial or the
final one; if we fix the initial condition, it is then not ensured that we
will find an acceptable final state (it is very improbable, indeed), and
\textit{vice versa}.

What we need is a model, at least an approximated one, which satisfies some
conditions both at the start and at the end. If it does not solve exactly
the PDEs, we can nevertheless take it as a zeroth order, perturbing to the
right version, even numerically.

\subsubsection{$r$ as a label}

The crucial observation is that, for an only matter Universe, the evolution
law results to be $A(r;t)=r\left[ 1+\frac{3}{2}H_{0}(r)t\right] ^{2/3}$. A
FLRW Universe with pure matter has analogously $a(t)=\left[ 1+\frac{3}{2}%
H_{0}t\right] ^{2/3}$. Hence, the inhomogeneous Universe has, at radius $r$,
a metric $A(r;t)=ra_{r}(t)$ s.t. $\frac{\dot{a}_{r}^{2}}{a_{r}^{2}}%
=H_{0}(r)^{2}a_{r}^{-3}$. If one considers only the spherical region until $r
$, and the total matter inside is regarded as if it were homogeneous, the
subsequent evolution law in $t$ depends on the label $r$ exactly as the LTB
solution.

This observation works exactly only for pure matter. We see this by setting $%
v:=0$ and $f:=1$. The conservations of matter and radiation respectively
yield $\rho _{M}\propto (A^{\prime 2})^{-1}$ and $\rho _{R}\propto
(A^{\prime 2})^{-4/3}$. By plugging these into the first Einstein equation,
one obtains%
\begin{equation}
\partial _{r}(\dot{A}^{2}A)=8\pi G\left( \rho _{M0}+\frac{\rho _{R0}}{\sqrt[3%
]{A^{\prime 2}}}\right) A_{0}^{\prime }A_{0}^{2}.
\end{equation}%
One can realize that this is exactly integrable for pure matter, while the
radiation returns a non trivial term. Henceforth, we should bear in mind
that that the `radius as label'~method is just an approximation : it works
fairly well near M-AM, when the universe is almost FLRW, and near today,
when the matter dominates, whereas it gets worse during the intermediate
period. With this \textit{caveat}, we attempt at writing%
\begin{equation}
A(r;t):=ra_{r}(t)\quad \text{s.t.}\quad \frac{\dot{a}_{r}^{2}}{H_{0}(r)^{2}}%
=\Omega _{M0}(r)a_{r}^{-1}+\Omega _{R0}(r)a_{r}^{-2}.
\end{equation}%
Since we consider here just the final components, $\Omega _{K0}=0$. They are
defined as usual, s.t. $\Omega _{M0}+\Omega _{R0}=1$. We can solve this by
means of exact integrations, as follows :%
\begin{gather}
\frac{da}{dt}=\frac{H_{0}}{a}\sqrt{\Omega _{M0}a+\Omega _{R0}}; \\
\Downarrow   \notag \\
\int_{t_{BB}}^{t}H_{0}dt=\int_{t_{BB}}^{t}\frac{ada}{\sqrt{\Omega
_{M0}a+\Omega _{R0}}}=[2\Omega _{M0}a\sqrt{\Omega _{M0}a+\Omega _{R0}}%
]_{t_{BB}}^{t}-2[\frac{2}{3}(\Omega _{M0}a+\Omega _{R0})^{3/2}]_{t_{BB}}^{t};
\\
\Downarrow   \notag \\
\frac{3}{2}H_{0}(t-t_{BB})=(\Omega _{M0}a-2\Omega _{R0})\sqrt{\Omega
_{M0}a+\Omega _{R0}}+2\Omega _{R0}^{3/2},
\end{gather}%
where we used the fact $a(t_{BB}):=0$. Moreover, by setting $a(0):=1$, one
obtains%
\begin{gather}
-\frac{3}{2}H_{0}t_{BB}=(\Omega _{M0}-2\Omega _{R0})\sqrt{\Omega
_{M0}+\Omega _{R0}}+2\Omega _{R0}^{3/2}=\Omega _{M0}+2\Omega _{R0}(\Omega
_{R0}^{1/2}-1); \\
\Downarrow   \notag \\
\frac{3}{2}H_{0}t=(\Omega _{M0}a-2\Omega _{R0})\sqrt{\Omega _{M0}a+\Omega
_{R0}}+(2\Omega _{R0}-\Omega _{M0}).
\end{gather}%
This is an exact evolution law $a=a(t)$, albeit implicitly expressed; the
explicit dependence can be obtained by exploiting Cardano's formula.

Next, we proceed and set the parameter of the real Universe. First of all,
the time singularity $t_{BB}(r)$ must be spatially homogeneous; since $t=0$
is today, we can call $T$ the age of the Universe, so that $-t_{BB}\equiv T$%
. Therefore, we set the fractal $\rho _{M0}:=\frac{D\Phi }{4\pi }r^{D-3}$.
The evolution law evaluated at $t=-T$ yields the last constraint,%
\begin{equation}
\frac{3}{2}H_{0}T=\Omega _{M0}+2\Omega _{R0}(\Omega
_{R0}^{1/2}-1)\Rightarrow 2(1-\Omega _{M0})^{3/2}+3\Omega _{M0}-2=T\sqrt{%
\frac{3}{2}DG\Phi }r^{\frac{D-3}{2}}\Omega _{M0}^{1/2}.
\end{equation}%
Notice that $\Omega _{M0}(r)$ can be expressed as the solution to an high
order algebraic equation, and this fixes also $H_{0}$, $\Omega _{R0}$ and $%
\rho _{R0}$.

It is difficult to solve exactly the algebraic equation of $\Omega _{M0}(r)$%
. Here, we confine ourselves to provide an approximated solution, relying on
the fact that $\Omega _{R0}\ll \Omega _{M0}$. Indeed, the evolution equation
at $-T$ becomes%
\begin{equation}
\frac{3}{2}H_{0}T=\Omega _{M0}+2\Omega _{R0}(\Omega _{R0}^{1/2}-1)\cong
\Omega _{M0}-2\Omega _{R0}\cong \Omega _{M0}.
\end{equation}%
Substituting $\rho _{M0}$, one reaches the following result :%
\begin{gather}
\frac{3}{2}H_{0}T\cong \frac{2}{3}DG\Phi r^{D-3}H_{0}^{-2}\Rightarrow
H_{0}(r)\cong \sqrt[3]{\frac{4DG\Phi }{9T}}r^{\frac{D}{3}-1}; \\
\Downarrow   \notag \\
\Omega _{M0}(r)\cong \frac{3}{2}T\sqrt[3]{\frac{4DG\Phi }{9T}}r^{\frac{D}{3}%
-1}=\sqrt[3]{\frac{3}{2}DG\Phi T^{2}}r^{\frac{D}{3}-1}; \\
\rho _{R0}(r)=\frac{3H_{0}^{2}}{8\pi G}-\rho _{M0}\cong \frac{1}{4\pi }\left[
\sqrt[3]{\frac{2D^{2}\Phi ^{2}}{3GT^{2}}}r^{1-\frac{D}{3}}-D\Phi \right]
r^{D-3},
\end{gather}%
where the numerical parameters $D$, $\Phi $ and $T$ can be deduced from
astronomical measurements.

Notwithstanding the fact that the above formul\ae\ are quite simple, this
model has a major drawback: the expansion is not homogeneous near $-T$,
because also the radiation is not homogeneous; in fact, it goes as%
\begin{equation}
\dot{a^{2}}\sim ^{-T}H_{0}^{2}\Omega _{R0}a^{-2}\Rightarrow a_{r}(t)\sim
\rho _{R0}(r)^{1/4}\sqrt{4\left( \frac{2}{3}\pi G\right) ^{1/2}(t+T)},
\end{equation}%
which expands faster for bigger $r$.

\subsubsection{Step functions}

An improvement can be achieved by admitting an evolution of the $\Omega $'s.
Indeed, we know that the matter and radiation densities do not change just
because of the expansion, but they also move through $r$, as it is described
by the PDEs. That is why the radiation can be homogeneous near $-T$ and
inhomogeneous at $t=0$: $\Omega _{R0}$ changes with time.

This fact can be roughly described inserting initial and final values, $%
H_{I},\Omega _{MI},\Omega _{RI}$ resp. $H_{F},\Omega _{MF},\Omega _{RF}$. In
other words, $H_{0}=H_{0}(t)$ is a step function that jumps from $H_{I}$ to $%
H_{F}$, and the same holds for the others. The jumps takes place at some
middle instant $t_{m}$, at which we assume all the changes to be
concentrated.

The evolution law can be written with differentials as%
\begin{eqnarray}
Hdt &=&\frac{ada}{\sqrt{\Omega _{M}a+\Omega _{R}}}=d\left[ 2\Omega _{M}a%
\sqrt{\Omega _{M}a\Omega _{R}}\right] -d\left[ \frac{4}{3}(\Omega
_{M}a+\Omega _{R})^{3/2}\right]  \notag \\
&=&\frac{2}{3}d[(\Omega _{M}a-2\Omega _{R})\sqrt{\Omega _{M}a+\Omega _{R}}].
\end{eqnarray}%
Setting $a(-T):=0$ and $a(0):=1$, it respectively holds that%
\begin{equation}
\left\{
\begin{array}{lc}
\frac{3}{2}H_{I}(t+T)=(\Omega _{MI}a-2\Omega _{RI})\sqrt{\Omega
_{MI}a+\Omega _{RI}}+2\Omega _{RI}^{3/2}, & -T\leq t\leq t_{m}; \\
&  \\
\frac{3}{2}H_{F}t=(\Omega _{MF}a-2\Omega _{RF})\sqrt{\Omega _{MF}a+\Omega
_{RF}}-(\Omega _{MF}-2\Omega _{RF}), & t_{m}\leq t\leq 0.%
\end{array}%
\right.
\end{equation}

Since the Einstein equations are of second order, it must be $a(t)\in
C^{1}(t_{m})$. Calling $a_{m}:=a(t_{m})$, we can write such request as%
\begin{equation}
\begin{cases}
\frac{1}{H_{I}}\left[ (\Omega _{MI}a_{m}-2\Omega _{RI})\sqrt{\Omega
_{MI}a_{m}+\Omega _{RI}}+2\Omega _{RI}^{3/2}\right] =\frac{3}{2}(t_{m}+T)=
\\
=\frac{1}{H_{F}}\left[ (\Omega _{MF}a_{m}-2\Omega _{RF})\sqrt{\Omega
_{MF}a_{m}+\Omega _{RF}}-(\Omega _{MF}-2\Omega _{RF})\right] +\frac{3}{2}T;
\\
\\
H_{I}^{2}(\Omega _{MI}a_{m}+\Omega _{RI})=\dot{a}%
_{m}^{2}a_{m}^{2}=H_{F}^{2}(\Omega _{MF}a_{m}+\Omega _{RF})%
\end{cases}%
\end{equation}%
Moreover, we can set the initial and final states as%
\begin{equation}
\begin{cases}
\Omega _{MI}(r)+\Omega _{RI}(r)=1=\Omega _{MF}(r)+\Omega _{RF}(r); \\
\\
H_{I}(r)^{2}\Omega _{RI}(r)=\frac{8}{3}\pi G\rho _{RI};\quad
H_{F}(r)^{2}\Omega _{MF}(r)=\frac{2}{3}DG\Phi r^{D-3}.%
\end{cases}%
\end{equation}%
These are overall 6 conditions involving 7 functions: $\Omega _{MI}(r)$, $%
\Omega _{RI}(r)$, $\Omega _{MF}(r)$, $\Omega _{RF}(r)$, $H_{I}(r)$, $H_{F}(r)
$ and $a_{m}(r)$ (which is equivalent to $t_{m}$).

In order to obtain the seventh condition, we recall that both the matter and
radiation densities depend on $v$, according to the following conservation
laws :%
\begin{equation}
\begin{cases}
\partial _{t}(V^{\prime }\rho _{M})=\partial _{r}(A^{2}v\rho _{M}); \\
\\
\partial _{t}\left( (V^{\prime 4/3}\rho _{R}\right) =\frac{4}{3}\sqrt[3]{%
V^{\prime }}\partial _{r}(A^{2}v\rho _{R}).%
\end{cases}%
\end{equation}%
Since here the $\rho $'s jump at $t_{m}$, all these derivatives have a Dirac
delta peak. For this reason, we can neglect the variations $\partial
_{t}(V^{\prime })$, $\partial _{r}(\rho _{M})$ and $\partial _{r}(\rho _{R})$%
, and take them approximately constant w.r.t. the jumps. Consequently, the
conservation laws become%
\begin{equation}
\begin{cases}
V^{\prime }\Delta (\rho _{M})\cong \rho _{M}\Delta (A^{2}v); \\
\\
(V^{\prime 4/3}\Delta \left( \rho _{R}\right) \cong \frac{4}{3}\sqrt[3]{%
V^{\prime }}\rho _{R}\Delta (A^{2}v);%
\end{cases}%
\Rightarrow \frac{4}{3}\Delta (\ln \rho _{M})\cong \frac{\Delta (A^{2}v)}{%
V^{\prime }}\cong \Delta (\ln \rho _{R}),
\end{equation}%
where the $\Delta $'s on the $\rho $'s are intended as $\Delta
(f):=\lim_{t\rightarrow t_{m}^{+}}f(t)-\lim_{t\rightarrow t_{m}^{-}}f(t)$.
Thus, one can rewrite%
\begin{equation}
\frac{4}{3}(\ln \rho _{MF}-\ln \rho _{MI})\cong \ln \rho _{RF}-\ln \rho
_{RI}\Leftrightarrow \left( \frac{\rho _{MF}}{\rho _{MI}}\right) ^{4/3}\cong
\frac{\rho _{RF}}{\rho _{RI}}.
\end{equation}%
Thence, one is able to determine completely the functions involved into the
model, with the remaining parameters being just numbers: $T$, $\rho _{RI}$, $%
D$ and $\Phi $.

In order to obtain a manageable set of algebraic equations, we enforce the
approximations $\rho _{RI}\gg \rho _{MI}$, $\rho _{MF}\gg \rho _{RF}$, and $%
a_{m}\gg 0$, then reaching the following results :%
\begin{gather}
\Omega _{RI}\cong 1,\quad \Omega _{MI}\cong 0,\quad \Omega _{MF}\cong
1,\quad \Omega _{RF}\cong 0; \\
\Downarrow   \notag \\
H_{I}(r)\cong \sqrt{\frac{8}{3}\pi G\rho _{RI}},\quad H_{F}(r)\cong \sqrt{%
\frac{8}{3}\pi G\rho _{MF}}=\sqrt{\frac{2}{3}DG\Phi }r^{\frac{D-3}{2}},
\end{gather}%
and%
\begin{eqnarray}
0 &\cong &\frac{1}{H_{I}}\left[ (\Omega _{MI}a_{m}-2\Omega _{RI})\sqrt{%
\Omega _{MI}a_{m}+\Omega _{RI}}+2\Omega _{RI}^{3/2}\right]   \notag \\
&=&\frac{1}{H_{F}}\left[ (\Omega _{MF}a_{m}-2\Omega _{RF})\sqrt{\Omega
_{MF}a_{m}+\Omega _{RF}}-(\Omega _{MF}-2\Omega _{RF})\right] +\frac{3}{2}%
T\cong \frac{a_{m}^{3/2}-1}{H_{F}}-\frac{3}{2}T,
\end{eqnarray}%
yielding%
\begin{equation}
a_{m}(r)\cong \left[ 1-\frac{3}{2}TH_{F}\right] ^{2/3}\cong \left[ 1-T\sqrt{%
\frac{3}{2}DG\Phi }r^{\frac{D-3}{2}}\right] ^{2/3}.
\end{equation}%
Then, we find from the other constraints that%
\begin{gather}
\frac{8}{3}\pi G\rho _{RI}\cong H_{I}^{2}(\Omega _{MI}a_{m}+\Omega
_{RI})=H_{F}^{2}(\Omega _{MF}a_{m}+\Omega _{RF})=\frac{8}{3}\pi G(\rho
_{MF}a_{m}+\rho _{RF}); \\
\Downarrow   \notag \\
\rho _{RF}(r)\cong \rho _{RI}-\rho _{MF}a_{m}\cong \rho _{RI}-\frac{D\Phi }{%
4\pi }r^{D-3}\left[ 1-T\sqrt{\frac{3}{2}DG\Phi }r^{\frac{D-3}{2}}\right]
^{2/3},
\end{gather}%
and%
\begin{equation}
\left( \frac{\rho _{MF}}{\rho _{MI}}\right) ^{4/3}\cong \frac{\rho _{RF}}{%
\rho _{RI}}\Rightarrow \rho _{MI}(r)\cong \rho _{MF}\left( \frac{\rho _{RI}}{%
\rho _{RF}}\right) ^{3/4}\cong \rho _{MF}\left( \frac{\rho _{RI}}{\rho
_{RI}-\rho _{MF}a_{m}}\right) ^{3/4}.
\end{equation}%
The four parameters of our simplified model can be empirically fixed, in
order to quantitatively\ compare the theoretical predictions with the
observational measurements. Following \cite{fract}, we evaluate $D\cong 1.2$%
, between the magnitude orders $L_{G}\cong 10^{5}ly$ and $L_{EG}\cong
3\times 10^{8}ly$. The fractal density can thus be obtained from the amount
of observed matter,%
\begin{gather}
\Phi L_{EG}^{D}\cong M(L_{EG})\cong \frac{4}{3}\pi L_{EG}^{3}\rho _{B0}=%
\frac{H_{0}^{2}}{2G}L_{EG}^{3}\Omega _{B0}; \\
\Downarrow   \notag \\
\Phi \cong \frac{H_{0}^{2}}{2G}L_{EG}^{3-D}\Omega _{B0}\cong 9.974\times
10^{24}kg/ly^{D},
\end{gather}%
where $H_{0}\cong 6.867\times 10^{-11}y^{-1}$ and $\Omega _{B0}\cong 0.044$
are the parameters of the Cosmological Concordance Model (CCM). Analogously,
we can evaluate the amount of radiation,%
\begin{gather}
\frac{H_{0}^{2}}{2G}L_{EG}^{3}\Omega _{R0}\cong \int_{0}^{L_{EG}}\rho
_{RF}4\pi r^{2}dr\cong \frac{4}{3}\pi L_{G}^{3}\rho
_{RI}+\int_{L_{G}}^{L_{EG}}\left[ 4\pi r^{2}\rho _{RI}-D\Phi r^{D-1}\left(
1-T\sqrt{\frac{3}{2}DG\Phi }r^{\frac{D-3}{2}}\right) ^{2/3}\right] dr  \notag
\\
=\frac{4}{3}\pi L_{EG}^{3}\rho _{RI}-\left[ \Phi r^{D}{}_{2}F_{1}\left( -%
\frac{2}{3},\frac{2D}{D-3};3\frac{D-1}{D-3};T\sqrt{\frac{3}{2}DG\Phi }r^{%
\frac{D-3}{2}}\right) \right] _{L_{G}}^{L_{EG}}; \\
\Downarrow   \notag \\
\rho _{RI}\cong \rho _{R0}+2.5\rho _{MF}(L_{EG}){}_{2}F_{1}\left( -\frac{2}{3%
},-\frac{4}{3};-\frac{1}{3};T\sqrt{1.8G\Phi }L_{EG}^{-0.9}\right)   \notag \\
-2.5\left( \frac{L_{G}}{L_{EG}}\right) ^{3}\rho
_{MF}(L_{G}){}_{2}F_{1}\left( -\frac{2}{3},-\frac{4}{3};-\frac{1}{3};T\sqrt{%
1.8G\Phi }L_{G}^{-0.9}\right) ,
\end{gather}%
where $\frac{8\pi G}{3H_{0}^{2}}\rho _{R0}:=\Omega _{R0}\cong 8.24\times
10^{-5}$ are CCM parameters again, and ${}_{2}F_{1}$ denotes the
hypergeometric function. The remaining parameter $T$ can be evaluated by
fitting further cosmological observations.

\section{Conclusions}

In this paper we have started a systematic development of the framework
focussed on the analysis of the consequences of fractal cosmology on the
evolution of the Universe. We have proposed a genesis of the cosmic fractal,
as well as a partition in epochs in Sec. 1.4, both suitable to obtain
quantitative results. Only the first epoch can consistently be described
with the usual FLRW solution; on the other hand, the LTB solution was
exploited for the second epoch in Sec. 3, and we proved in Sec. 2 that an
even more general Lema\^{\i}tre solution is necessary for the third epoch,
because of general restrictions arising from the momentum conservation in
the LTB metric.

Of course, our calculations admit further improvements, for instance
provided by a more precise solution to the evolution equations of the Lema%
\^{\i}tre model, as discussed in Sec. 4. After our analysis, we may
reasonably wonder that a more detailed analysis would describe the fall of
the matter fractal onto itself, thus providing self-consistency and
stability within fractal cosmology, while the homogeneous FLRW would just be
an unstable solution. Future works might also improve the description of the
second epoch, e.g. implementing the transformation law of matter into
radiation.

It is worth pointing out here that the whole theoretical framework dealing
with LTB and Lema\^{\i}tre models provides a smooth approximation to the
actual fractal dynamics. Indeed, a more realistic model for fractal
cosmology should make use of distributional General Relativity, which is a
quite formidable task, or at least of a first order perturbative
approximation towards the anisotropic distribution. These latter
perturbative methods, applied to an LTB or Lema\^{\i}tre background, should
expectedly provide some amount of effective dark matter, due to retarded
potentials \cite{re}, \cite{re2}. Since the fractal approach is able to
explain dark energy phenomena \cite{Cosmai:2018nvx}, it is conceivable that
a combined framework will be able to overcome many of the drawbacks of the
Cosmological Concordance Model.

Finally, we would like to remark that a deeper quantitative analysis of the
LTB metric is of potential relevance also in other frameworks, such as the
IR-completion of gravity \cite{piazza1, piazza2}, or within the attempts to
explain the tension of the Hubble parameter \cite{fanizza}.

\subsection*{Acknowledgments}

\medskip We would like to thank Alessio Notari, for his careful analysis and
constructive criticism in an early stage of the project, as well as Federico Piazza, for drawing our attention to his work on the LTB model. Also, we
acknowledge useful correspondence with Francesco Sylos Labini, Luigi Tedesco
and Giuseppe Fanizza. Last but not least, we are especially grateful to
Luciano Pietronero, for sharing his insightful view on the matter-antimatter
recombination epoch.


\end{document}